\documentclass[aps,prd,twocolumn,preprintnumbers,showpacs,superscriptaddress,nofootinbib,amsmath,amssymb,floats,floatfix,showkeys,notitLepage1989,longbibliography]{revtex4-1}
\usepackage{graphicx}
\usepackage{color}
\usepackage[colorlinks=true,linkcolor=blue,citecolor=magenta,urlcolor=blue]{hyperref}
\usepackage{appendix}
\usepackage{float}
\usepackage[normalem]{ulem}
\usepackage{bm}
\usepackage{bigints}
\usepackage{array,booktabs}
\usepackage{color,amsmath,amssymb,epsfig,graphicx}
\usepackage{bbm}
\usepackage{mathrsfs}
\usepackage{tikz}
\usepackage{pgfplots}
\usepgfplotslibrary{groupplots}
\usepgfplotslibrary{groupplots}
\usepgfplotslibrary{fillbetween}
\pgfplotsset{compat=1.18}
\usepackage{caption}
\usepackage{ragged2e}
\usepackage{url}
\usepackage{soul}
\usepackage{cancel}
\usepackage{xcolor}
\usepackage{comment}
\usepackage{float}
\definecolor{njlblue}{RGB}{31,119,180}
\definecolor{gupred}{RGB}{214,39,40}

\allowdisplaybreaks

\begin{document}

\title{Minimal-Length Deformations of Chiral Quark Matter in a 2+1-Flavor NJL Model}

\author{Sameer Ahmad Mir}
\email{sameerphst@gmail.com}
\affiliation{Canadian Quantum Research Center, 460 Doyle Ave 106, Kelowna, BC V1Y 0C2, Canada}
%\affiliation{Department of Physics, Jamia Millia Islamia, New Delhi, 110025, India}
\affiliation{Department of Computer Sciences, Asian School of Business, Noida, Uttar Pradesh, 201303, India}

\begin{abstract} 
In this paper, a minimal-length inspired deformation of hot and dense chiral quark matter is formulated within a \(2+1\)-flavor Nambu-Jona-Lasinio framework. The deformation is based on a complete Jacobi-consistent three-dimensional quadratic GUP algebra and modifies the microscopic phase-space density of states while leaving the quasiparticle spectrum unchanged, leading to a thermodynamically closed mean-field theory with coupled light and strange gap equations. The analysis shows that ultraviolet phase-space suppression weakens dynamical chiral symmetry breaking, lowers the light and strange constituent masses, and shifts the light-sector chiral transition boundary, including its crossover and critical endpoint toward lower temperature and chemical potential. 
The numerical results are obtained from a single self-consistent solution of the coupled gap equations, and explicit endpoint comparisons show that the temperature-dependent and chemical-potential dependent results are mutually consistent.  
The result is a systematic effective-model study of how minimal-length inspired ultraviolet deformations reorganize coupled chiral dynamics and bulk quark-matter thermodynamics.
\end{abstract}

\date{\today}

\maketitle

\twocolumngrid

\section{Introduction}

Understanding strongly interacting matter under extreme conditions remains one
of the central problems of modern high-energy and nuclear physics. The thermal
behavior of quantum chromodynamics (QCD) matter governs the interpretation of
heavy-ion collisions, constrains the structure of the QCD phase diagram, and is
closely connected with the microscopic origin of chiral symmetry breaking and
its restoration in hot and dense environments~\cite{Koch:1995vp,Banks:1979yr,Csaki:2021jax}.
Recent studies of hadron production, finite baryon-size effects, and
thermodynamically consistent hadronic equations of state further emphasize the
need for a reliable medium description across the hadronic side of the QCD phase
diagram~\cite{Mir:2023wkm,Mir:2024wlo,Mir:2025qqv}. 
The finite-temperature behavior of kaons and antikaons in isospin-asymmetric dense resonance matter has also been
studied~\cite{Kaur:2024cfm}. The in-medium properties of the $\phi$ meson in dense resonance matter at finite temperature have also been investigated~\cite{Kaur:2025kjk}. 
The most challenging region for these questions occurs at low temperature,
moderate baryon density, and strong coupling. In this region, perturbative QCD
is not applicable, while lattice QCD faces the sign problem at finite baryon
chemical potential. Effective theories that preserve the relevant symmetry
structure of QCD are therefore indispensable~\cite{Buballa:2003qv}. 
Among such theories, the Nambu-Jona-Lasinio (NJL) model occupies a central
place. Although it does not incorporate confinement, it provides a transparent
and self-consistent description of spontaneous chiral symmetry breaking,
dynamical constituent-mass generation, and chiral restoration at finite
temperature and density. In this sense, the NJL framework isolates the chiral
sector of QCD in a form that is analytically tractable and thermodynamically
consistent, making it one of the standard effective tools for studying the
nonperturbative structure of quark matter~\cite{Nambu:1961tp,Nambu:1961fr,Buballa:2003qv}.
Because its thermodynamics is governed by explicit momentum-space integrals,
the model is especially well suited for examining how microscopic modifications
of phase-space structure affect bulk thermodynamic observables.

A conceptually distinct but increasingly active direction comes from
quantum-gravity phenomenology. Several approaches to Planckian short distance
physics suggest that the usual short-distance structure of quantum theory may
be modified by a minimal length, a deformed Heisenberg algebra, or a nontrivial
momentum-space geometry~\cite{Maggiore:1993rv,Das:2008kaa}. 
In effective descriptions, these ideas are commonly represented through
generalized uncertainty principles (GUPs), modified commutators, modified
dispersion relations, or deformed phase-space measures
\cite{Hossenfelder:2012jw,Kempf:1994su,Ali:2009zq,Faizal:2015Conseq,Faizal:2015Lifshitz,Faizal:2016SUSYGUP}.
One of their robust qualitative consequences is the suppression of ultraviolet
phase-space support. From the perspective of many-body theory, this is
important because condensates, equations of state, susceptibilities, and phase
boundaries are all controlled by momentum integrals and may therefore be
sensitive to such deformations~\cite{Klevansky:1992qe}. 
The possibility that minimal-length effects may influence strongly interacting
matter has already attracted attention
\cite{Salam:1993xy,Sijacki:1990xp,Brindejonc:1995pr,Sivaram:1979mg,Naggar:2013foi,Nozari:2015qoi}. 
Recent phenomenological work has also considered quantum-gravity motivated
deformations in heavy-ion observables, including elliptic flow, which further
motivates the study of GUP effects in QCD matter~\cite{Mir:2026cbr}. 
GUP inspired corrections have been studied in simplified analyses of
quark gluon plasma (QGP) thermodynamics, and deformed momentum measures have also
been examined in the SU(2) NJL model, particularly for constraining the
size of the deformation parameter from vacuum observables
\cite{Naggar:2013foi,Nozari:2015qoi}. These studies show that the connection
between QCD matter and minimal-length phenomenology is physically motivated.
What is still needed is a finite-temperature, finite-density, self-consistent
chiral formalism in which the deformation enters directly at the level of the
grand potential and propagates through the gap equations into the
thermodynamics of quark matter. Such a formulation allows one to ask how
minimal-length motivated ultraviolet deformations modify coupled chiral
dynamics in an effective QCD model. 
The present work should therefore be understood as a systematic mean-field
study of the ultraviolet sensitivity of chiral quark matter, not as a claim
that dynamical quantum gravity acts inside QCD matter. The GUP enters only
through a deformation of microscopic phase-space state counting. The
deformation parameter is treated as a phenomenological ultraviolet-sensitivity
parameter within the NJL model. The calculation asks how the coupled
light-strange chiral mean-field system responds when the density of
high-momentum quark states is suppressed by a minimal-length inspired
phase-space measure. This interpretation is essential because the canonical
Planckian value of a fundamental GUP parameter would be far too small to
produce observable effects in heavy-ion or compact-star QCD matter. The
parameter values used below should therefore be viewed as phenomenological
sensitivity tests rather than direct estimates of quantum-gravity effects.

The NJL model is particularly useful for this purpose because the relevant
observables are obtained self-consistently. The constituent quark masses, scalar
condensates, pressure, and susceptibilities are all determined from the same
stationary grand potential. A deformation of the momentum-space measure
therefore does not act as an isolated correction to a single thermodynamic
quantity; rather, it enters at the microscopic level and propagates through the
entire mean-field system. This has two advantages. First, it identifies which
parts of QCD matter phenomenology are most sensitive to ultraviolet
phase-space suppression. Second, it provides a natural way to discuss
phenomenological bounds: even if the deformation is small, one may still ask
how large it can be before the chiral condensate, the constituent mass, or the
phase structure deviates from standard NJL phenomenology
\cite{Nozari:2015qoi}. 
Here we formulate a generalized uncertainty principle extension of the
\(2+1\)-flavor NJL model at mean-field level. The deformation is introduced
through an isotropic modification of the momentum-space measure, while the
quasiparticle dispersion relation is kept in its standard NJL form. This
isolates the GUP effect in the density of momentum states and preserves the
stationarity structure of the thermodynamic potential. Within this formulation
we derive the deformed grand potential, obtain the corresponding coupled gap
equations, and analyze the leading consequences for dynamical mass generation
and chiral restoration. We show that, in the present implementation, the
deformation suppresses the ultraviolet support of the scalar gap integrals,
weakens dynamical chiral symmetry breaking, and shifts the light-sector
pseudocritical line toward lower temperature and lower quark chemical
potential. These trends already follow from the analytical structure of the gap
equations and are then confirmed by the numerical solutions. 
More broadly, the formalism presented here connects two active areas: the study
of hot and dense QCD through chiral effective models and the study of
minimal-length effects through phenomenological modifications of microscopic
phase space. It provides a self-consistent route for incorporating
quantum-gravity motivated phase-space deformations into an effective theory of
quark matter while preserving thermodynamic closure. In this sense, the present
model should be viewed as a study of ultraviolet sensitivity in the chiral
sector and as a starting point for more realistic extensions, most naturally to
a PNJL+GUP formulation in which deconfinement-like physics is included together
with chiral dynamics~\cite{Fukushima:2003fw,Ratti:2006wg,Fukushima:2008wg}. A broader
analysis of phase structure may also benefit from recent functional and
holographic approaches to the QCD critical end point
\cite{Mir:2026fort,Burikham:2017bkn,Erlich:2005qh,Sakai:2004cn}.

\section{Standard 2+1-flavor NJL model at mean field}
\label{sec:njl_meanfield}

Before introducing the generalized uncertainty principle deformation, we first
present the undeformed \(2+1\)-flavor NJL model in an explicit mean-field form.
This section fixes the flavor structure of the model, gives the light- and
strange-sector constituent masses, and writes the corresponding
finite-temperature and finite-density grand potential. In contrast to the
two-flavor case, the \(2+1\)-flavor theory contains an additional
flavor-mixing interaction associated with the
Kobayashi-Maskawa-'t~Hooft determinant term. This interaction couples the
light and strange condensates and is essential for discussing \(M_l\) and
\(M_s\) in one self-consistent mean-field system
\cite{Klevansky:1992qe,Hatsuda:1994pi,Rehberg:1995kh}. We begin with the
standard \(SU(3)_f\) NJL Lagrangian,
\begin{multline}
\mathcal{L}
=
\bar q\,(i\gamma^\mu\partial_\mu-\hat m)\,q
+
G\sum_{a=0}^{8}
\left[
(\bar q\lambda_a q)^2
+
(\bar q i\gamma_5\lambda_a q)^2
\right]
-\\
K
\left[
\det \bar q(1+\gamma_5)q
+
\det \bar q(1-\gamma_5)q
\right],
\label{eq:njl_3fl_lagrangian}
\end{multline}
where \(q=(u,d,s)^T\) is the quark field in flavor space,
\(\hat m=\mathrm{diag}(m_u,m_d,m_s)\) is the current-mass matrix,
\(\lambda_a\) are the Gell-Mann matrices in flavor space together with
\(\lambda_0=\sqrt{2/3}\,\mathbf{1}\), \(G\) is the four-fermion
scalar-pseudoscalar coupling, and \(K\) is the six-fermion flavor-mixing
coupling associated with the \(U_A(1)\)-breaking determinant interaction.
Throughout this work we impose isospin symmetry in the light sector,
\(
m_u=m_d\equiv m_l
%\label{eq:isospin_mass}
\)
and therefore treat the light condensates symmetrically. 
The relevant mean-field order parameters are the flavor-diagonal scalar
condensates
\(
\phi_u \equiv \langle \bar u u\rangle,\:
\phi_d \equiv \langle \bar d d\rangle,\:
\phi_s \equiv \langle \bar s s\rangle.
%\label{eq:phi_f_def}
\)
Under isospin symmetry we set
\(
\phi_u=\phi_d\equiv \phi_l,
%\label{eq:phi_l_def}
\)
so that the independent mean-field variables are \(\phi_l\) and \(\phi_s\).
For homogeneous, parity-even matter, the pseudoscalar condensates vanish.
Expanding the interaction terms about the scalar mean fields gives the
constituent masses
\begin{equation}
M_l
=
m_l-4G\phi_l+2K\,\phi_l\phi_s,
\label{eq:Ml_def}
\end{equation}
and
\begin{equation}
M_s
=
m_s-4G\phi_s+2K\,\phi_l^2.
\label{eq:Ms_def}
\end{equation}
Thus the light and strange constituent masses are coupled nonlinearly through
the determinant interaction. This coupling is the reason why the strange-quark
effective mass cannot be introduced consistently in a purely two-flavor NJL
model. At mean-field level, the condensate contribution to the thermodynamic
potential is
\begin{equation}
\Omega_{\mathrm{cond}}
=
2G\left(2\phi_l^2+\phi_s^2\right)
-
4K\,\phi_l^2\phi_s.
\label{eq:Omega_cond}
\end{equation}
The quasiparticle energies in the light and strange sectors are
\(
E_{l,s}(p)=\sqrt{p^2+M_{l,s}^2}.\:
\)
In the present analysis, we use a common quark chemical potential,
\(
\mu_u=\mu_d=\mu_s\equiv \mu,
%\label{eq:common_mu}
\)
so that the same \(\mu\) fixes the thermodynamic environment of both sectors. 
The reduction to two independent condensates in the common-$\mu$
analysis requires isospin symmetry both in the current masses,
$m_u=m_d$, and in the light-flavor chemical potentials,
$\mu_u=\mu_d$. When flavor-dependent chemical potentials are
introduced, $\mu_u$ and $\mu_d$ need not coincide. Consequently, the
Fermi-Dirac occupation factors and the medium contributions to the
$u$- and $d$-sector gap equations are generally different, and
$\phi_u$ and $\phi_d$ must be retained as independent mean fields.
Accordingly, the zero-temperature analysis of charge-neutral matter in
beta equilibrium is carried out with three independent condensates,
$(\phi_u,\phi_d,\phi_s)$. The notation $2+1$ denotes the current-mass
pattern $m_u=m_d\neq m_s$; it does not imply
$\phi_u=\phi_d$ in an isospin-asymmetric medium. 
% The same mean-field equations can be extended directly to
% flavor-dependent chemical potentials.
Prior to ultraviolet regularization, the mean-field grand-potential
density of the undeformed \(2+1\)-flavor NJL model is
\begin{widetext}
\begin{align}
\Omega_{\rm NJL}^{2+1}(\phi_l,\phi_s,T,\mu)
&=
2G\left(2\phi_l^2+\phi_s^2\right)
-
4K\,\phi_l^2\phi_s
\nonumber\\
&\quad
-
2N_c
\int \frac{d^3p}{(2\pi)^3}
\Bigg\{
2\left[
E_l
+
T\ln\!\left(1+e^{-(E_l-\mu)/T}\right)
+
T\ln\!\left(1+e^{-(E_l+\mu)/T}\right)
\right]
\nonumber\\
&\hspace{2.8cm}
+
\left[
E_s
+
T\ln\!\left(1+e^{-(E_s-\mu)/T}\right)
+
T\ln\!\left(1+e^{-(E_s+\mu)/T}\right)
\right]
\Bigg\}.
\label{eq:omega_njl_2p1}
\end{align}
\end{widetext}
Here, the factor of \(2\) multiplying the light-sector contribution accounts
for the degeneracy of the \(u\) and \(d\) flavors. Because the NJL model is
nonrenormalizable, an explicit ultraviolet regularization scheme must be
specified. Throughout this work, we impose a common sharp three-momentum
cutoff on the full quasiparticle contribution, including both the vacuum and
thermal terms. The same momentum interval is retained in the undeformed and
GUP-deformed formulations, so that the difference between them originates
solely from the deformed density of states factor and the corresponding
self-consistent mean-field solution. The regularized undeformed
grand-potential density is
\begin{widetext}
\begin{align}
\Omega_{\mathrm{NJL}}^{2+1,(\Lambda)}
(\phi_l,\phi_s,T,\mu)
={}&
2G\left(2\phi_l^2+\phi_s^2\right)
-4K\phi_l^2\phi_s
\nonumber\\
&-2N_c
\int_{|\boldsymbol p|<\Lambda}
\frac{d^3p}{(2\pi)^3}
\Bigg\{
2\Big[
E_l
+T\ln\!\left(1+e^{-(E_l-\mu)/T}\right)
+T\ln\!\left(1+e^{-(E_l+\mu)/T}\right)
\Big]
\nonumber\\
&\hspace{4.0cm}
+\Big[
E_s
+T\ln\!\left(1+e^{-(E_s-\mu)/T}\right)
+T\ln\!\left(1+e^{-(E_s+\mu)/T}\right)
\Big]
\Bigg\}.
\label{eq:njl_potential_cutoff}
\end{align}
\end{widetext}
All stationary equations and thermodynamic quantities used below are
derived from this fixed-\(\Lambda\) potential. To avoid unnecessary
notation, the superscript \((\Lambda)\) is suppressed after this
definition, but the common cutoff remains understood throughout. 
The equilibrium values of
\(\phi_l\) and \(\phi_s\) follow from the stationarity conditions
\begin{equation}
\frac{\partial \Omega_{\rm NJL}^{2+1}}{\partial \phi_l}=0,
\qquad
\frac{\partial \Omega_{\rm NJL}^{2+1}}{\partial \phi_s}=0.
\label{eq:stationarity_2p1}
\end{equation}
Equivalently, the condensates can be written as momentum integrals over the
corresponding quasiparticle modes,
\begin{equation}
\phi_l=
-2N_c
\int_{|\boldsymbol p|<\Lambda}
\frac{d^3p}{(2\pi)^3}
\frac{M_l}{E_l}
\left[
1-f_l^-(E_l)-f_l^+(E_l)
\right],
\label{eq:phil_integral}
\end{equation}
\begin{equation}
\phi_s=
-2N_c
\int_{|\boldsymbol p|<\Lambda}
\frac{d^3p}{(2\pi)^3}
\frac{M_s}{E_s}
\left[
1-f_s^-(E_s)-f_s^+(E_s)
\right].
\label{eq:phis_integral}
\end{equation}
where
\begin{equation}
f_l^\mp(E_l)=\frac{1}{e^{(E_l\mp\mu)/T}+1},
\quad
f_s^\mp(E_s)=\frac{1}{e^{(E_s\mp\mu)/T}+1}.
\label{eq:ff_pm}
\end{equation}
where \(E_{l}=\sqrt{p^2+M_{l}^2}\) and \(E_{s}=\sqrt{p^2+M_{s}^2}\) is the energy of the light and strange quasiparticles, respectively. 
Using spherical symmetry, these expressions reduce to
\begin{equation}
\phi_l
=
-\frac{N_c}{\pi^2}
\int_0^\Lambda dp\,
p^2
\frac{M_l}{E_l}
\Big[
1-f_l^-(E_l)-f_l^+(E_l)
\Big],
\label{eq:phil_radial}
\end{equation}
\begin{equation}
\phi_s
=
-\frac{N_c}{\pi^2}
\int_0^\Lambda dp\,
p^2
\frac{M_s}{E_s}
\Big[
1-f_s^-(E_s)-f_s^+(E_s)
\Big].
\label{eq:phis_radial}
\end{equation}
Eqs.~\eqref{eq:Ml_def}, \eqref{eq:Ms_def}, \eqref{eq:phil_radial}, and
\eqref{eq:phis_radial} define the closed coupled nonlinear gap system for the
light and strange sectors. This undeformed mean-field system provides the
reference point for introducing the GUP-modified phase-space measure.

\section{GUP-deformed phase space and the 2+1-flavor NJL+GUP grand potential}
\label{sec:gup_measure}

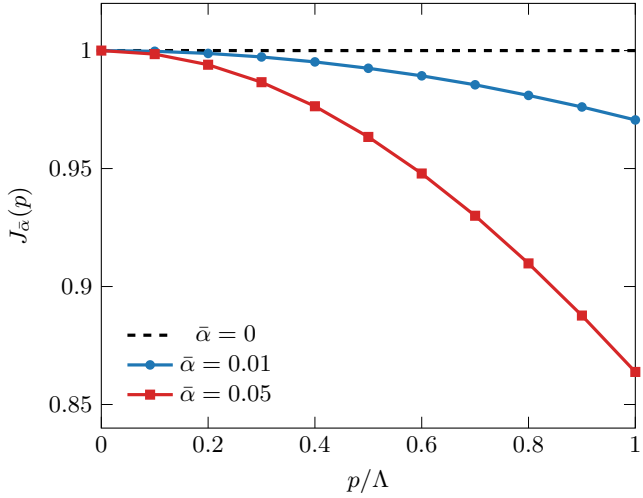
\begin{figure}[htb]
\centering
\begin{tikzpicture}
\begin{axis}[
    width=\linewidth,
    height=7.2cm,
    xlabel={$p/\Lambda$},
    ylabel={$J_{\bar{\alpha}}(p)$},
    xmin=0, xmax=1,
    ymin=0.84, ymax=1.02,
    legend style={
        draw=none,
        fill=none,
        at={(0.03,0.03)},
        anchor=south west,
        font=\small
    },
    tick label style={font=},%\small},
    label style={font=},%\small},
    axis line style={black},
    tick style={black}
]

\addplot[
    very thick,
    dashed,
    color=black,
    mark=none
]
table[row sep=\\]{
x J \\
0.0 1.0000 \\
0.1 1.0000 \\
0.2 1.0000 \\
0.3 1.0000 \\
0.4 1.0000 \\
0.5 1.0000 \\
0.6 1.0000 \\
0.7 1.0000 \\
0.8 1.0000 \\
0.9 1.0000 \\
1.0 1.0000 \\
};
\addlegendentry{\(\bar{\alpha}=0\)}

\addplot[
    very thick,
    color=njlblue,
    mark=*,
    mark size=1.2pt,
    mark options={fill=njlblue}
]
table[row sep=\\]{
x J \\
0.0 1.0000 \\
0.1 0.9997 \\
0.2 0.9988 \\
0.3 0.9973 \\
0.4 0.9952 \\
0.5 0.9925 \\
0.6 0.9893 \\
0.7 0.9855 \\
0.8 0.9810 \\
0.9 0.9761 \\
1.0 0.9706 \\
};
\addlegendentry{\(\bar{\alpha}=0.01\)}

\addplot[
    very thick,
    color=gupred,
    mark=square*,
    mark size=1.3pt,
    mark options={fill=gupred}
]
table[row sep=\\]{
x J \\
0.0 1.0000 \\
0.1 0.9985 \\
0.2 0.9940 \\
0.3 0.9866 \\
0.4 0.9764 \\
0.5 0.9634 \\
0.6 0.9479 \\
0.7 0.9300 \\
0.8 0.9098 \\
0.9 0.8877 \\
1.0 0.8638 \\
};
\addlegendentry{\(\bar{\alpha}=0.05\)}
\end{axis}
\end{tikzpicture}
\caption{\justifying Deformation factor \(J_{\bar{\alpha}}(p)=(1+\bar{\alpha}(p/\Lambda)^2)^{-3}\)
over the NJL cutoff interval. The reference value \(\bar{\alpha}=0.01\)
produces only a mild reduction of the high-momentum phase-space weight, while
\(\bar{\alpha}=0.05\) is included only to make the ultraviolet momentum
dependence more visible.}
\label{fig:Jacobian}
\end{figure}

Having established the undeformed \(2+1\)-flavor NJL framework, we
now specify the implementation of the generalized uncertainty
principle deformation. Deformed momentum-space measures have
previously been considered in NJL phenomenology, notably in the
\(SU(2)\) study of Nozari et al.~\cite{Nozari:2015qoi}. That study,
however, used a compact momentum-space formulation that differs from
the quadratic GUP phase-space measure employed here. We consider the
\(\beta'=0\) sector of the rotationally invariant
Kempf-Mangano-Mann algebra~\cite{Kempf:1994su}. The corresponding
semiclassical invariant phase-space measure and momentum-space density
of states are obtained from the analysis of Chang
et al.~\cite{Chang:2001bm}. In three spatial dimensions, the resulting
phase-space weight is
\(
J_\alpha(p)=(1+\alpha p^2)^{-3}.
\)
This factor multiplies the momentum integrals, whereas the
quasiparticle dispersion relation remains in its standard NJL form.
The GUP modification is therefore confined to the momentum-space
density of states, while rotational invariance and the coupled
mean-field stationarity conditions are preserved. The resulting
light-strange NJL+GUP framework suppresses the ultraviolet
contributions to both the light- and strange-sector gap equations.
The isotropic quadratic GUP adopted in the present analysis is
the $\beta'=0$ sector of the rotationally invariant
Kempf-Mangano-Mann algebra~\cite{Kempf:1994su}. In three
spatial dimensions, its Jacobi-consistent operator form is
\begin{equation}
\begin{aligned}
 [X_i,P_j] &= i\hbar F(P^2)\delta_{ij},\\
 [P_i,P_j] &= 0,\\
 [X_i,X_j] &= 2i\hbar\alpha
 \left(P_iX_j-P_jX_i\right),
\end{aligned}
\label{eq.gup_complete_algebra}
\end{equation}
where \(F(P^2)\equiv 1+\alpha P^2 ,\) $P^2=\sum_{k=1}^{3}P_kP_k$, $\alpha>0$ has dimensions
of inverse momentum squared, and $i,j=1,2,3$. The final
commutator in Eq.~\eqref{eq.gup_complete_algebra} is required
for compatibility with the multidimensional Jacobi identity.
For the nontrivial $(X_i,X_j,P_k)$ identity, one obtains
\begin{multline}
[X_i,[X_j,P_k]]+[X_j,[P_k,X_i]]\\
 =2\hbar^2\alpha F(P^2)
 \left(\delta_{ik}P_j-\delta_{jk}P_i\right),
\\
\hspace{-6cm}[P_k,[X_i,X_j]]
\\ 
=-2\hbar^2\alpha F(P^2)
 \left(\delta_{ik}P_j-\delta_{jk}P_i\right),
\end{multline}
so that the complete Jacobi sum vanishes identically. The same
algebra is realized in momentum space through
$P_i\psi(\boldsymbol p)=p_i\psi(\boldsymbol p)$ and
$X_i\psi(\boldsymbol p)
 =i\hbar(1+\alpha p^2)\partial\psi/\partial p_i$. 
The Robertson inequality then gives, for each spatial
coordinate,
\begin{equation}
\Delta X_i\,\Delta P_i
\geq
\frac{\hbar}{2}
\left[
1+\alpha\sum_{k=1}^{3}
\left(
(\Delta P_k)^2+\langle P_k\rangle^2
\right)
\right].
\label{eq.gup_uncertainty_3d}
\end{equation}
For $\langle\boldsymbol P\rangle=0$, minimization with respect
to the momentum uncertainty gives the absolute coordinate-wise
minimum
\begin{equation}
(\Delta X_i)_{\min}=\hbar\sqrt{\alpha}.
\label{eq.minimum_length}
\end{equation} 
In the semiclassical limit, the complete operator algebra
becomes \cite{Benczik:2002tt}
\begin{equation}
\begin{aligned}
 \{x_i,p_j\} &= f(p^2)\delta_{ij},\\
 \{p_i,p_j\} &= 0,\\
 \{x_i,x_j\} &= 2\alpha
 \left(p_ix_j-p_jx_i\right),
\end{aligned}
\label{eq.complete_poisson_algebra}
\end{equation}
where \(f(p^2)\equiv 1+\alpha p^2 \). Because the coordinates are noncommutative, the invariant
phase-space density must be obtained from the full
six-dimensional Poisson tensor rather than from its mixed
position-momentum block alone. We therefore define
$z^A=(x_1,x_2,x_3,p_1,p_2,p_3)$, the Poisson tensor and its
determinant are given as \cite{CannasDaSilva2001}
\begin{equation}
\begin{aligned}
\Theta^{AB}
\equiv \{z^A,z^B\}
=
\begin{pmatrix}
 A & f\,\mathbf{1}_3\\
 -f\,\mathbf{1}_3 & 0
\end{pmatrix},
\\
A_{ij}
=2\alpha(p_ix_j-p_jx_i),\,\,\,
\det\Theta
&=f^6(p^2).
\end{aligned}
\label{eq.poisson_tensor}
\end{equation}
Let $\omega_{AB}=(\Theta^{-1})_{AB}$ denote the corresponding
symplectic form. Its Liouville density is
\begin{equation}
\rho_\alpha(p)
=\sqrt{\det\omega}
=\frac{1}{\sqrt{\det\Theta}}
=\frac{1}{f^3(p^2)}.
\end{equation}
It also satisfies
\(
\partial_A\!\left(\rho_\alpha\Theta^{AB}\right)=0,
\)
which implies
\(\partial_A(\rho_\alpha\dot z^A)=0\) for Hamiltonian evolution
and hence shows that the weighted phase-space volume is conserved
under Hamiltonian flow. The corresponding invariant phase-space
element is therefore~\cite{Chang:2001bm}
\begin{equation}
d\Gamma_\alpha
=
\frac{d^3x\,d^3p}{(2\pi\hbar)^3}\,
\rho_\alpha(p)
=
\frac{d^3x\,d^3p}{(2\pi\hbar)^3}
\frac{1}{(1+\alpha p^2)^3}.
\label{eq.invariant_phase_space}
\end{equation}
In the natural units used in the thermodynamic calculation,
$\hbar=1$. Dividing Eq.~\eqref{eq.invariant_phase_space} by
the spatial volume therefore gives the momentum-space
replacement
\begin{equation}
\frac{d^3p}{(2\pi)^3}
\longrightarrow
\frac{d^3p}{(2\pi)^3}J_\alpha(p),
\,\,\,
J_\alpha(p)=\frac{1}{(1+\alpha p^2)^3}.
\label{eq.gup_measure}
\end{equation}
Thus, the momentum-space factor used in the grand potential
is the three-dimensional semiclassical Liouville density
associated with the complete Jacobi-consistent algebra in
Eq.~\eqref{eq.gup_complete_algebra}. The exponent three is
fixed by the number of spatial dimensions. This
thermodynamic density-of-states factor should not be confused
with the single factor $f^{-1}$ appearing in the
one-particle momentum-space scalar product of the operator
representation. 
The positivity of \(\alpha\) guarantees ultraviolet suppression, since \(J_\alpha(p)<1\) for \(p>0\), while the undeformed NJL model is recovered smoothly in the limit \(\alpha\to0\). 
This choice satisfies
\begin{multline}
\lim_{\alpha\to0}\mathcal{J}_\alpha(p)=1,
\quad
\frac{d\mathcal{J}_\alpha}{dp}
=
-\frac{6\alpha p}{(1+\alpha p^2)^4}<0
\\
(\alpha>0,\ p>0),
\label{eq:jacobian_properties_2p1}
\end{multline}
so the deformation is manifestly ultraviolet suppressing. For \(\alpha p^2\ll1\),
\begin{equation}
\mathcal{J}_\alpha(p)
=
1-3\alpha p^2+\mathcal{O}(\alpha^2),
\label{eq:jacobian_small_alpha_2p1}
\end{equation}
which makes explicit that the leading correction increasingly suppresses
the high-momentum contribution as \(p\) increases. Fig.~\ref{fig:Jacobian}
shows the deformation factor \(J_{\bar{\alpha}}(p)\) as a function of
\(p/\Lambda\). For the reference value \(\bar{\alpha}=0.01\), the
high-momentum phase-space weight is reduced only modestly throughout the
cutoff interval \(0\leq p\leq\Lambda\). The larger values included for
comparison make the ultraviolet suppression more pronounced.

The GUP-deformed potential is obtained by replacing the momentum
measure in the fixed-cutoff undeformed potential,
Eq.~\eqref{eq:njl_potential_cutoff}, according to Eq.~\eqref{eq.gup_measure}
while retaining the same integration domain
\(|\boldsymbol p|<\Lambda\). The resulting regularized
grand-potential density is
\begin{widetext}
\begin{align}
\Omega_{\mathrm{NJL+GUP}}^{2+1,(\Lambda)}
(\phi_l,\phi_s,T,\mu)
={}&
2G\left(2\phi_l^2+\phi_s^2\right)
-4K\phi_l^2\phi_s
\nonumber\\
&-2N_c
\int_{|\boldsymbol p|<\Lambda}
\frac{d^3p}{(2\pi)^3}
J_\alpha(p)
\Bigg\{
2\Big[
E_l
+T\ln\!\left(1+e^{-(E_l-\mu)/T}\right)
+T\ln\!\left(1+e^{-(E_l+\mu)/T}\right)
\Big]
\nonumber\\
&\hspace{4.0cm}
+\Big[
E_s
+T\ln\!\left(1+e^{-(E_s-\mu)/T}\right)
+T\ln\!\left(1+e^{-(E_s+\mu)/T}\right)
\Big]
\Bigg\}.
\label{eq:gup_potential_cutoff}
\end{align}
\end{widetext}
Using
\begin{equation}
\frac{d^3p}{(2\pi)^3}
=
\frac{p^2\,dp}{2\pi^2},
\end{equation}
the same potential takes the radial form
\begin{widetext}
\begin{align}
\Omega_{\mathrm{NJL+GUP}}^{2+1,(\Lambda)}
(\phi_l,\phi_s,T,\mu)
={}&
2G\left(2\phi_l^2+\phi_s^2\right)
-4K\phi_l^2\phi_s
\nonumber\\
&-\frac{N_c}{\pi^2}
\int_0^\Lambda dp\,
p^2J_\alpha(p)
\Bigg\{
2\Big[
E_l
+T\ln\!\left(1+e^{-(E_l-\mu)/T}\right)
+T\ln\!\left(1+e^{-(E_l+\mu)/T}\right)
\Big]
\nonumber\\
&\hspace{3.6cm}
+\Big[
E_s
+T\ln\!\left(1+e^{-(E_s-\mu)/T}\right)
+T\ln\!\left(1+e^{-(E_s+\mu)/T}\right)
\Big]
\Bigg\}.
\label{eq:gup_potential_radial}
\end{align}
\end{widetext}
Because the cutoff \(\Lambda\) and the integration domain are held
fixed, \(J_\alpha(p)\to1\) uniformly on
\(0\leq p\leq\Lambda\) as \(\alpha\to0\). It therefore follows
directly that, at fixed condensates \(\phi_l\) and \(\phi_s\),
\begin{equation}
\lim_{\alpha\to0}
\Omega_{\mathrm{NJL+GUP}}^{2+1,(\Lambda)}
(\phi_l,\phi_s,T,\mu)
=
\Omega_{\mathrm{NJL}}^{2+1,(\Lambda)}
(\phi_l,\phi_s,T,\mu).
\label{eq:fixed_cutoff_undeformed_limit}
\end{equation}
Thus, the undeformed NJL theory is recovered at fixed finite
\(\Lambda\), with the same ultraviolet convention on both sides of
the limit. No momentum integral extending to infinity is used in the
numerical, perturbative, or thermodynamic analysis below. As in the
undeformed case, the superscript \((\Lambda)\) is henceforth
suppressed.
The equilibrium condensates in the deformed theory again follow from
stationarity,
\begin{equation}
\frac{\partial \Omega_{\rm NJL+GUP}^{2+1}}{\partial \phi_l}=0,
\qquad
\frac{\partial \Omega_{\rm NJL+GUP}^{2+1}}{\partial \phi_s}=0.
\label{eq:stationarity_gup_2p1}
\end{equation}
Equivalently, the deformed condensates can be written as
\begin{equation}
\phi_l^{(\alpha)}
=
-2N_c
\int_{|\boldsymbol p|<\Lambda} \frac{d^3p}{(2\pi)^3}
\frac{1}{(1+\alpha p^2)^3}
\frac{M_l}{E_l}
\Big[
1-f_l^-(E_l)-f_l^+(E_l)
\Big],
\label{eq:phil_gup_cart}
\end{equation}
\begin{equation}
\phi_s^{(\alpha)}
=
-2N_c
\int_{|\boldsymbol p|<\Lambda} \frac{d^3p}{(2\pi)^3}
\frac{1}{(1+\alpha p^2)^3}
\frac{M_s}{E_s}
\Big[
1-f_s^-(E_s)-f_s^+(E_s)
\Big],
\label{eq:phis_gup_cart}
\end{equation}
or, after angular integration,
\begin{equation}
\phi_l^{(\alpha)}
=
-\frac{N_c}{\pi^2}
\int_0^\Lambda dp\,
\frac{p^2}{(1+\alpha p^2)^3}
\frac{M_l}{E_l}
\Big[
1-f_l^-(E_l)-f_l^+(E_l)
\Big],
\label{eq:phil_gup_rad}
\end{equation}
\begin{equation}
\phi_s^{(\alpha)}
=
-\frac{N_c}{\pi^2}
\int_0^\Lambda dp\,
\frac{p^2}{(1+\alpha p^2)^3}
\frac{M_s}{E_s}
\Big[
1-f_s^-(E_s)-f_s^+(E_s)
\Big].
\label{eq:phis_gup_rad}
\end{equation}
The coupled constituent masses are then
\begin{equation}
M_l
=
m_l-4G\phi_l^{(\alpha)}+2K\,\phi_l^{(\alpha)}\phi_s^{(\alpha)},
\label{eq:Ml_gup}
\end{equation}
\begin{equation}
M_s
=
m_s-4G\phi_s^{(\alpha)}+2K\,\big(\phi_l^{(\alpha)}\big)^2.
\label{eq:Ms_gup}
\end{equation}
Eqs.~\eqref{eq:phil_gup_rad}-\eqref{eq:Ms_gup} define the coupled
light-strange NJL+GUP gap system. A small-\(\alpha\) expansion gives useful
analytic guidance. Using
\begin{equation}
\frac{1}{(1+\alpha p^2)^3}
=
1-3\alpha p^2+\mathcal{O}(\alpha^2),
\label{eq:jacobian_expand_again_2p1}
\end{equation}
one obtains, at fixed \(\phi_l\) and \(\phi_s\),
\begin{widetext}
\begin{align}
\Omega_{\rm NJL+GUP}^{2+1}
&=
\Omega_{\rm NJL}^{2+1}
+
3\alpha\,\frac{N_c}{\pi^2}
\int_0^\Lambda dp\;p^4
\Bigg\{
2\left[
E_l
+
T\ln\!\left(1+e^{-(E_l-\mu)/T}\right)
+
T\ln\!\left(1+e^{-(E_l+\mu)/T}\right)
\right]
\nonumber\\
&\hspace{2.9cm}
+
\left[
E_s
+
T\ln\!\left(1+e^{-(E_s-\mu)/T}\right)
+
T\ln\!\left(1+e^{-(E_s+\mu)/T}\right)
\right]
\Bigg\}
+
\mathcal{O}(\alpha^2).
\label{eq:omega_small_alpha_2p1}
\end{align}
\end{widetext}
At fixed condensates, the positive-\(\alpha\) correction raises the grand
potential and therefore lowers the corresponding pressure. 
Along any differentiable stationary branch at fixed \(T\) and \(\mu\), the
implicit deformation dependence of the equilibrium condensates does not
contribute at first order to the stationary grand potential. Indeed,
\begin{equation}
\frac{d\Omega^\star}{d\alpha}
=
\left.
\frac{\partial\Omega}{\partial\alpha}
\right|_{\phi_l^\star,\phi_s^\star}
+
\sum_{f=l,s}
\left.
\frac{\partial\Omega}{\partial\phi_f}
\right|_{\star}
\frac{d\phi_f^\star}{d\alpha}
=
\left.
\frac{\partial\Omega}{\partial\alpha}
\right|_{\star},
\end{equation}
because the stationarity conditions imply
\(\left.\partial\Omega/\partial\phi_l\right|_{\star}
=\left.\partial\Omega/\partial\phi_s\right|_{\star}=0\).
Consequently, Eq.~\eqref{eq:omega_small_alpha_2p1} also fixes the leading positive-\(\alpha\) shift of
the stationary grand potential \(\Omega^\star\), and hence the leading
negative shift of the unnormalised stationary pressure
\(P^\star=-\Omega^\star\). 
Since the same
momentum-space weight multiplies the light and strange scalar integrals, both
sectors undergo ultraviolet phase-space suppression, although their numerical
responses need not be identical. The small-\(\alpha\) expansion therefore shows
that high-momentum modes are suppressed simultaneously in the light and strange
scalar channels.

\section{Results and Discussion}
\label{sec:analytical_results}

We now discuss the main analytical trends and numerical results of the
\(2+1\)-flavor NJL+GUP framework in a unified way. In contrast to the
two-flavor case, the equilibrium state is determined by two coupled scalar
condensates, \(\phi_l\) and \(\phi_s\), and by the corresponding constituent
masses \(M_l\) and \(M_s\). The GUP deformation therefore acts on two coupled
gap channels, with the light and strange sectors linked through the
Kobayashi-Maskawa-'t~Hooft flavor-mixing interaction. This section first
shows how the deformed phase-space measure modifies the coupled gap equations
and then derives the leading small-\(\alpha\) trends analytically. We then give
numerical results for the temperature and chemical-potential dependence of the
light and strange effective masses, together with the resulting pseudocritical
boundary. 
To facilitate numerical reproducibility, we specify the NJL model
parameters and the numerical procedure used to obtain the results shown
below. Throughout the numerical analysis, we employ the standard
\(2+1\)-flavor NJL parameterization of Rehberg, Klevansky, and
H\"ufner~\cite{Rehberg:1995kh}, which reproduces the vacuum pseudoscalar
observables of the undeformed theory. The corresponding parameter values
are listed in Table~\ref{tab:njl_parameters}, while the numerical
conditions used for each figure are summarized in
Table~\ref{tab:numerical_protocol}. We work in natural units and use the common sharp three-momentum cutoff defined in Eqs.~\eqref{eq:njl_potential_cutoff} and
\eqref{eq:gup_potential_cutoff} for all vacuum and thermal
contributions in both theories. We impose isospin symmetry in the
common-\(\mu\) light sector and solve the coupled gap equations
self-consistently at each \((T,\mu)\) point.

The dimensionless deformation parameter used in the numerical analysis is
\begin{equation}
\bar{\alpha}\equiv \alpha\Lambda^2.
\end{equation}
For the Rehberg-Klevansky-H\"ufner cutoff
\(\Lambda=0.6023~{\rm GeV}\), the reference value
\(\bar{\alpha}=0.01\) corresponds to
\begin{equation}
\alpha=\frac{0.01}{\Lambda^2}
=
2.7566\times10^{-2}~{\rm GeV}^{-2}.
\end{equation}
This smaller value is used to display the direction and magnitude of the
ultraviolet response while keeping the deformation perturbative at the cutoff
momentum. The earlier illustrative value \(\bar{\alpha}=0.25\) corresponds to
\(\alpha=0.689~{\rm GeV}^{-2}\) and produces a much stronger suppression of
the scalar gap integrals. It is therefore not used as the reference
deformation value in the numerical tables.

\begin{figure}[htb]
\centering
\begin{tikzpicture}
\begin{axis}[
    width=\linewidth,
    height=7.2cm,
    xlabel={$T\,[\mathrm{MeV}]$},
    ylabel={$M_l\,[\mathrm{MeV}]$},
    xmin=0, xmax=250,
    ymin=0, ymax=400,
    legend style={draw=none, fill=none, at={(0.03,0.27)}, anchor=north west, font=\small},
    tick label style={font=},%\small},
    label style={font=},%\small},
    axis line style={black},
    tick style={black}
]
\addplot[
    very thick,
    dashed,
    color=njlblue,
    mark=*,
    mark size=1.4pt,
    mark options={fill=njlblue}
] table[row sep=\\] {
T   Ml \\
0   367.648 \\
20  367.648 \\
40  367.640 \\
60  367.300 \\
80  365.110 \\
100 358.527 \\
120 344.728 \\
140 320.550 \\
150 303.280 \\
160 281.570 \\
170 254.363 \\
180 220.298 \\
190 178.295 \\
200 131.038 \\
210 90.433  \\
220 64.687  \\
230 49.570  \\
240 40.157  \\
250 33.856  \\
};
\addlegendentry{\textcolor{njlblue}{NJL}}
\addplot[
    very thick,
    color=gupred,
    mark=square*,
    mark size=1.6pt,
    mark options={fill=gupred}
] table[row sep=\\] {
T   Ml \\
0   349.766 \\
20  349.766 \\
40  349.754 \\
60  349.316 \\
80  346.704 \\
100 339.169 \\
120 323.691 \\
140 296.767 \\
150 277.523 \\
160 253.262 \\
170 222.786 \\
180 184.925 \\
190 140.708 \\
200 98.877  \\
210 70.042  \\
220 52.858  \\
230 42.289  \\
240 35.319  \\
250 30.432  \\
};
\addlegendentry{\textcolor{gupred}{NJL+GUP}}
\end{axis}
\end{tikzpicture}
\caption{\justifying Temperature dependence of the light-quark effective mass at \(\mu=0\) in the
undeformed NJL model and in the GUP-deformed case with
\(\bar{\alpha}=0.01\). The positive-\(\alpha\) deformation reduces the light
constituent mass throughout the chirally broken and crossover regions.}
\label{fig:MlT}
\end{figure}
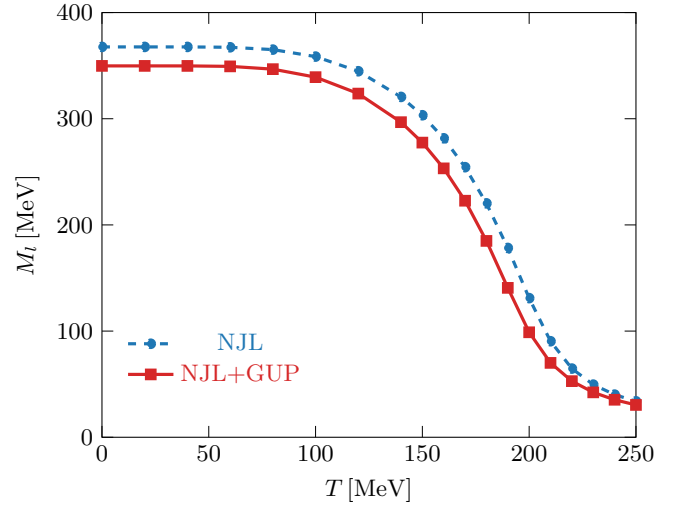

\begin{figure}[htb]
\centering
\begin{tikzpicture}
\begin{axis}[
    width=\linewidth,
    height=7.2cm,
    xlabel={$T\,[\mathrm{MeV}]$},
    ylabel={$M_s\,[\mathrm{MeV}]$},
    xmin=0,
    xmax=250,
    ymin=350,
    ymax=570,
    xtick={0,50,100,150,200,250},
    ytick={350,400,450,500,550},
    minor tick num=1,
    legend style={
        draw=none,
        fill=none,
        at={(0.03,0.27)},
        anchor=north west,
        font=\small,
        legend cell align=left
    },
    tick label style={font=},%\small},
    label style={font=},%\small},
    axis line style={black},
    tick style={black},
    clip marker paths=true
]

% Undeformed NJL solution
\addplot[
    very thick,
    dashed,
    color=njlblue,
    mark=*,
    mark repeat=2,
    mark size=1.5pt,
    mark options={
        solid,
        fill=njlblue,
        draw=njlblue
    }
] table[
    x=T,
    y=Ms,
    row sep=\\
] {
T   Ms \\
0   549.479 \\
10  549.479 \\
20  549.479 \\
30  549.479 \\
40  549.475 \\
50  549.444 \\
60  549.313 \\
70  548.952 \\
80  548.182 \\
90  546.790 \\
100 544.552 \\
110 541.240 \\
120 536.630 \\
130 530.497 \\
140 522.614 \\
150 512.749 \\
160 500.670 \\
170 486.172 \\
180 469.211 \\
190 450.320 \\
200 431.657 \\
210 416.655 \\
220 405.405 \\
230 395.822 \\
240 386.821 \\
250 378.052 \\
};
\addlegendentry{NJL}

% GUP-deformed solution: \bar{\alpha}=0.01
\addplot[
    very thick,
    solid,
    color=gupred,
    mark=square*,
    mark repeat=2,
    mark size=1.7pt,
    mark options={
        solid,
        fill=gupred,
        draw=gupred
    }
] table[
    x=T,
    y=Ms,
    row sep=\\
] {
T   Ms \\
0   534.616 \\
10  534.616 \\
20  534.616 \\
30  534.615 \\
40  534.610 \\
50  534.568 \\
60  534.406 \\
70  533.980 \\
80  533.096 \\
90  531.534 \\
100 529.060 \\
110 525.443 \\
120 520.452 \\
130 513.860 \\
140 505.439 \\
150 494.963 \\
160 482.231 \\
170 467.150 \\
180 450.016 \\
190 432.294 \\
200 417.044 \\
210 405.503 \\
220 395.958 \\
230 387.120 \\
240 378.531 \\
250 370.064 \\
};
\addlegendentry{NJL+GUP}

\end{axis}
\end{tikzpicture}

\caption{\justifying
Temperature dependence of the strange-quark constituent mass
$M_s$ at $\mu=0$ in the undeformed NJL model and in the
GUP-deformed model with $\bar{\alpha}=0.01$. Every plotted point
is obtained from the same self-consistent solution of the coupled
light- and strange-sector gap equations, Eqs.~\eqref{eq:Ml_gup}--
\eqref{eq:Ms_gup}, using the common three-momentum cutoff
$0\leq p\leq\Lambda$ and the parameter set listed in Table~\ref{tab:njl_parameters}.
At $T=50~\mathrm{MeV}$, the solutions are
$M_s^{\mathrm{NJL}}=549.444~\mathrm{MeV}$ and
$M_s^{\mathrm{NJL+GUP}}=534.568~\mathrm{MeV}$, exactly reproducing
the $\mu=0$ endpoints of the fixed-$T$ curves in Fig.~\ref{fig:Mmu}.
For an additional high-temperature check, at
$T=200~\mathrm{MeV}$ the corresponding values are
$431.657~\mathrm{MeV}$ and $417.044~\mathrm{MeV}$, while at
$T=250~\mathrm{MeV}$ they are $378.052~\mathrm{MeV}$ and
$370.064~\mathrm{MeV}$. The positive-$\alpha$ phase-space suppression therefore reduces the strange constituent mass throughout the displayed temperature interval,
although the reduction is smaller than in the light sector because of
the larger strange-quark current mass.}
\label{fig:MsT}
\end{figure}
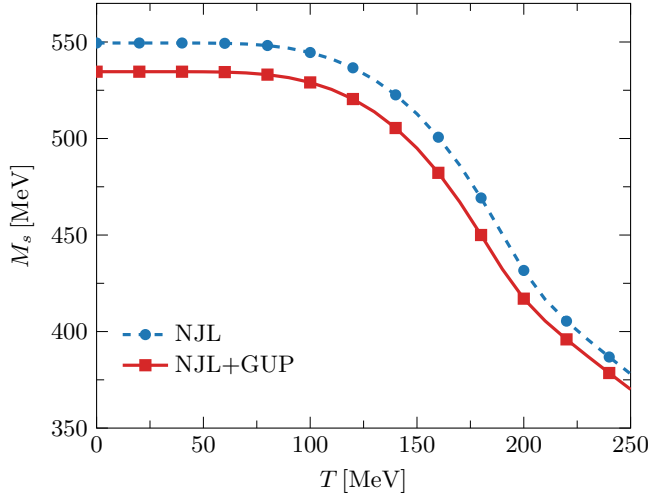

\begin{figure*}[htb]
\centering
\begin{tikzpicture}

\pgfplotsset{
    branchcurve/.style={
        very thick,
        smooth,
        tension=0.35,
        mark=none
    },
    transitioncurve/.style={
        very thick,
        densely dotted,
        mark=none
    },
    datapointsblue/.style={
        only marks,
        color=njlblue,
        mark=*,
        mark size=1.05pt,
        mark options={fill=njlblue, draw=njlblue}
    },
    datapointsred/.style={
        only marks,
        color=gupred,
        mark=square*,
        mark size=1.15pt,
        mark options={fill=gupred, draw=gupred}
    }
}

\begin{groupplot}[
    group style={group size=2 by 1, horizontal sep=1.8cm},
    width=8.2cm,
    height=7.2cm,
    xlabel={$\mu\,[\mathrm{MeV}]$},
    xmin=0, xmax=500,
    tick label style={font=},%\small},
    label style={font=},%\small},
    title style={font=},%\small},
    legend style={draw=none, fill=none, font=\small},
    axis line style={black},
    tick style={black},
    grid=none,
    clip=false
]

% -------------------------------
% (a) Light sector
% -------------------------------
\nextgroupplot[
    ylabel={$M_l\,[\mathrm{MeV}]$},
    ymin=0, ymax=400,
    title={(a) Light sector},
    legend pos=south west
]

% NJL: broken branch
\addplot[
    branchcurve,
    dashed,
    color=njlblue
] table[row sep=\\] {
mu   Ml \\
0    367.571 \\
50   367.530 \\
100  367.359 \\
150  366.869 \\
200  365.498 \\
250  361.517 \\
300  348.104 \\
320  332.887 \\
330  316.848 \\
};
\addlegendentry{\textcolor{njlblue}{NJL}}

% NJL: rapid branch change
\addplot[
    transitioncurve,
    dashed,
    color=njlblue,
    forget plot
] table[row sep=\\] {
mu   Ml \\
330  316.848 \\
340  70.478  \\
};

% NJL: restored branch
\addplot[
    branchcurve,
    dashed,
    color=njlblue,
    forget plot
] table[row sep=\\] {
mu   Ml \\
340  70.478  \\
350  50.990  \\
400  21.513  \\
450  13.014  \\
500  8.885   \\
};

% NJL: actual computed points
\addplot[
    datapointsblue,
    forget plot
] table[row sep=\\] {
mu   Ml \\
0    367.571 \\
50   367.530 \\
100  367.359 \\
150  366.869 \\
200  365.498 \\
250  361.517 \\
300  348.104 \\
320  332.887 \\
330  316.848 \\
340  70.478  \\
350  50.990  \\
400  21.513  \\
450  13.014  \\
500  8.885   \\
};

% NJL+GUP: broken branch
\addplot[
    branchcurve,
    color=gupred
] table[row sep=\\] {
mu   Ml \\
0    349.661 \\
50   349.604 \\
100  349.368 \\
150  348.693 \\
200  346.788 \\
250  341.115 \\
300  319.749 \\
320  284.278 \\
};
\addlegendentry{\textcolor{gupred}{NJL+GUP}}

% NJL+GUP: rapid branch change
\addplot[
    transitioncurve,
    color=gupred,
    forget plot
] table[row sep=\\] {
mu   Ml \\
320  284.278 \\
330  73.488  \\
};

% NJL+GUP: restored branch
\addplot[
    branchcurve,
    color=gupred,
    forget plot
] table[row sep=\\] {
mu   Ml \\
330  73.488  \\
340  53.153  \\
350  41.962  \\
400  19.959  \\
450  12.468  \\
500  8.728   \\
};

% NJL+GUP: actual computed points
\addplot[
    datapointsred,
    forget plot
] table[row sep=\\] {
mu   Ml \\
0    349.661 \\
50   349.604 \\
100  349.368 \\
150  348.693 \\
200  346.788 \\
250  341.115 \\
300  319.749 \\
320  284.278 \\
330  73.488  \\
340  53.153  \\
350  41.962  \\
400  19.959  \\
450  12.468  \\
500  8.728   \\
};

% -------------------------------
% (b) Strange sector
% -------------------------------
\nextgroupplot[
    ylabel={$M_s\,[\mathrm{MeV}]$},
    ymin=250, ymax=600,
    title={(b) Strange sector},
    legend pos=south west
]

% NJL: high-mass branch
\addplot[
    branchcurve,
    dashed,
    color=njlblue
] table[row sep=\\] {
mu   Ms \\
0    549.444 \\
50   549.424 \\
100  549.345 \\
150  549.118 \\
200  548.484 \\
250  546.655 \\
300  540.618 \\
320  534.002 \\
330  527.302 \\
};
\addlegendentry{\textcolor{njlblue}{NJL}}

% NJL: light-sector induced branch change
\addplot[
    transitioncurve,
    dashed,
    color=njlblue,
    forget plot
] table[row sep=\\] {
mu   Ms \\
330  527.302 \\
340  462.393 \\
};

% NJL: intermediate strange branch
\addplot[
    branchcurve,
    dashed,
    color=njlblue,
    forget plot
] table[row sep=\\] {
mu   Ms \\
340  462.393 \\
350  459.927 \\
400  449.887 \\
450  416.332 \\
};

% NJL: dense strange-branch onset
\addplot[
    transitioncurve,
    dashed,
    color=njlblue,
    forget plot
] table[row sep=\\] {
mu   Ms \\
450  416.332 \\
500  272.997 \\
};

% NJL: actual computed points
\addplot[
    datapointsblue,
    forget plot
] table[row sep=\\] {
mu   Ms \\
0    549.444 \\
50   549.424 \\
100  549.345 \\
150  549.118 \\
200  548.484 \\
250  546.655 \\
300  540.618 \\
320  534.002 \\
330  527.302 \\
340  462.393 \\
350  459.927 \\
400  449.887 \\
450  416.332 \\
500  272.997 \\
};

% NJL+GUP: high-mass branch
\addplot[
    branchcurve,
    color=gupred
] table[row sep=\\] {
mu   Ms \\
0    534.568 \\
50   534.542 \\
100  534.437 \\
150  534.133 \\
200  533.280 \\
250  530.763 \\
300  521.598 \\
320  507.510 \\
};
\addlegendentry{\textcolor{gupred}{NJL+GUP}}

% NJL+GUP: light-sector induced branch change
\addplot[
    transitioncurve,
    color=gupred,
    forget plot
] table[row sep=\\] {
mu   Ms \\
320  507.510 \\
330  455.062 \\
};

% NJL+GUP: intermediate strange branch
\addplot[
    branchcurve,
    color=gupred,
    forget plot
] table[row sep=\\] {
mu   Ms \\
330  455.062 \\
340  452.494 \\
350  450.809 \\
400  439.918 \\
450  399.715 \\
};

% NJL+GUP: dense strange-branch onset
\addplot[
    transitioncurve,
    color=gupred,
    forget plot
] table[row sep=\\] {
mu   Ms \\
450  399.715 \\
500  263.744 \\
};

% NJL+GUP: actual computed points
\addplot[
    datapointsred,
    forget plot
] table[row sep=\\] {
mu   Ms \\
0    534.568 \\
50   534.542 \\
100  534.437 \\
150  534.133 \\
200  533.280 \\
250  530.763 \\
300  521.598 \\
320  507.510 \\
330  455.062 \\
340  452.494 \\
350  450.809 \\
400  439.918 \\
450  399.715 \\
500  263.744 \\
};

\end{groupplot}
\end{tikzpicture}

\caption{\justifying
Chemical-potential dependence of the light- and strange-quark effective
masses at fixed temperature \(T=50~\mathrm{MeV}\) in the undeformed NJL model
and in the GUP-deformed model with \(\bar{\alpha}=0.01\). The \(\mu=0\)
endpoints are \(M_l=367.571~\mathrm{MeV}\) and
\(M_s=549.444~\mathrm{MeV}\) for NJL, and
\(M_l=349.661~\mathrm{MeV}\) and \(M_s=534.568~\mathrm{MeV}\) for NJL+GUP,
in agreement with the \(T=50~\mathrm{MeV}\) values obtained from
Figs.~\ref{fig:MlT} and \ref{fig:MsT}. At this temperature, the
discontinuous light-sector branch change occurs at
\(\mu_{\chi}^{\mathrm{NJL}}=335.26~\mathrm{MeV}\) in the undeformed model
and at
\(\mu_{\chi}^{\mathrm{NJL+GUP}}=323.22~\mathrm{MeV}\) in the
GUP-deformed model, accompanied by a corresponding strange-sector response
through the Kobayashi-Maskawa-'t~Hooft flavor-mixing interaction. The curves
are shown separately on each resolved branch, while the dotted connectors
indicate rapid branch changes. 
%; no spline interpolation across a smooth crossover is implied. 
The symbols denote the self-consistent gap-equation
solutions at the displayed chemical potentials.}
\label{fig:Mmu}
\end{figure*}
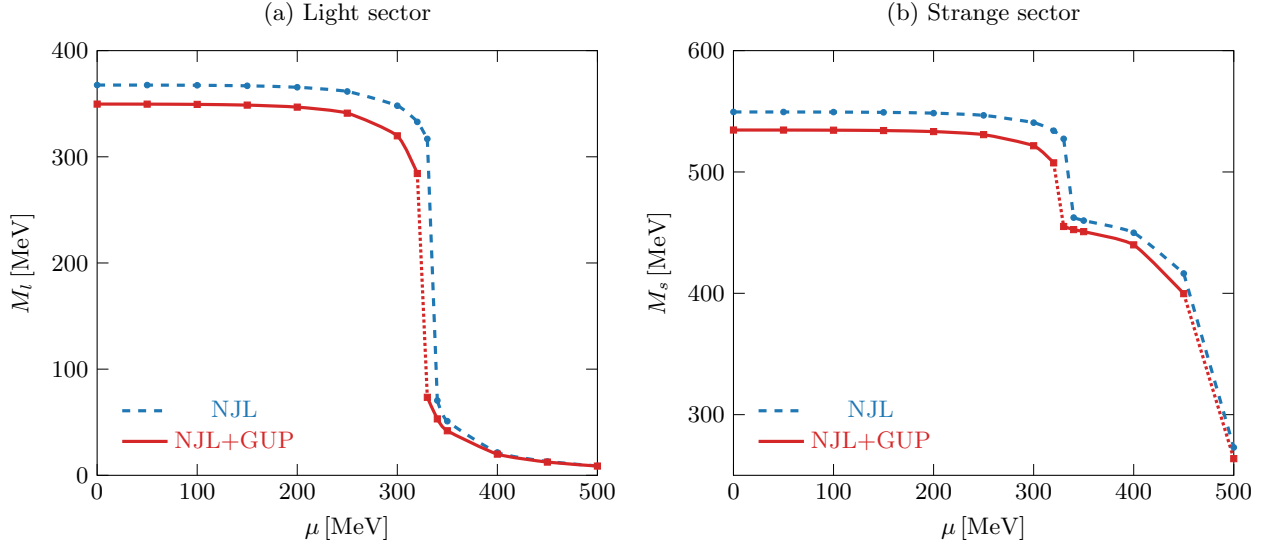

As indicated in Table~\ref{tab:numerical_protocol}, all
temperature-dependent mass, pressure, and susceptibility curves are evaluated
at \(\mu=0\), whereas all chemical-potential dependent mass and pressure curves
are evaluated at fixed temperature \(T=50~\mathrm{MeV}\). 
The crossover segment of the light-sector chiral boundary is obtained
from the peak of the thermal response
\(\chi_{T,l}=-\partial M_l/\partial T\), equivalently from the inflection
point of \(M_l(T,\mu)\). At lower temperature, the first-order
coexistence segment is determined from equality of the competing
broken- and restored-branch stationary grand-potential minima. The
meeting point of the crossover and first-order segments defines the
critical endpoint. The
coupled gap equations were solved self-consistently at every \((T,\mu)\) point
by iterating the light and strange condensates until both constituent masses
changed by less than \(10^{-8}\,\mathrm{GeV}\) between successive iterations.
The momentum integrals were evaluated using adaptive quadrature and
independently repeated with Gauss-Legendre quadrature using
\(N_p=800\), \(1200\), and \(1600\) nodes. The resulting constituent
masses and thermodynamic observables remain unchanged at the quoted
precision as \(N_p\) is increased. For the perturbative analysis, the Jacobian of the coupled gap system in Eq.~\eqref{eq:kernel-source-system}  was also evaluated on every retained stable
branch. Its diagonal principal minors and determinant
$\mathcal D$ remained positive away from the critical
singularity, confirming that the inverse response matrix is
regular in the regions where the first-order expansion is
used. 
As a direct endpoint comparison among the figures, the \(\mu=0\)
endpoint of the fixed-\(T\) curve in Fig.~\ref{fig:Mmu} was compared
with the corresponding values at \(T=50~\mathrm{MeV}\) from the
temperature-dependent curves in Figs.~\ref{fig:MlT} and
\ref{fig:MsT}. The same self-consistent solution gives
\begin{equation}
\begin{array}{c|cc}
& \bar{\alpha}=0 & \bar{\alpha}=0.01 \\
\hline
M_l(T=50~\mathrm{MeV},\mu=0)\,[\mathrm{MeV}] & 367.571 & 349.661 \\
M_s(T=50~\mathrm{MeV},\mu=0)\,[\mathrm{MeV}] & 549.444 & 534.568
\end{array}
\end{equation}
so that Figs.~\ref{fig:MlT}, \ref{fig:MsT}, and \ref{fig:Mmu} are obtained
from one common numerical solution. The pressure curves are normalized by
subtracting the vacuum value of the stationary grand potential separately in
each theory,
\begin{equation}
P_{\rm sub}(T,\mu,\alpha)
=
-\left[
\Omega^\star(T,\mu,\alpha)
-
\Omega^\star(0,0,\alpha)
\right],
\label{eq.40}
\end{equation}
where \(\Omega^\star\) denotes the grand potential evaluated at the
self-consistent stationary solution.
\begin{table*}[htb]
\centering
\caption{\justifying
Input \(2+1\)-flavor NJL parameter set used in the numerical analysis. The
dimensionless couplings are quoted together with their equivalent values in
GeV units.}
\label{tab:njl_parameters}
\begin{tabular}{ccccccc}
\hline
\(m_l\) & \(m_s\) & \(\Lambda\) & \(G\Lambda^2\) & \(K\Lambda^5\) & \(G\) & \(K\) \\
\hline
\(5.5~\mathrm{MeV}\) & \(140.7~\mathrm{MeV}\) & \(602.3~\mathrm{MeV}\) & \(1.835\) & \(12.36\) & \(5.06~\mathrm{GeV}^{-2}\) & \(155.9~\mathrm{GeV}^{-5}\) \\
\hline
\end{tabular}
\end{table*}
\begin{table*}[htb]
\centering
\caption{\justifying Numerical inputs used for the figures. The dimensionless
deformation strength is defined as \(\bar{\alpha}\equiv\alpha\Lambda^2\). The
numerical analysis uses \(\bar{\alpha}=0.01\), while additional values are
included only to illustrate the momentum-space deformation factor.}
\label{tab:numerical_protocol}
\begin{tabular}{lll}
\hline
Quantity / Figure & Fixed conditions & Numerical choice \\
\hline
Jacobian \(J_{\bar{\alpha}}(p)\) & -- & \(\bar{\alpha}=0,\;0.01,\;0.05\) \\
\(M_l(T),\,M_s(T)\) & \(\mu=0\) & \(\bar{\alpha}=0.01\) \\
\(M_l(\mu),\,M_s(\mu)\) & fixed \(T\) & \(T=50~{\rm MeV},\;\bar{\alpha}=0.01\) \\
\(P(T)\) & \(\mu=0\) & \(\bar{\alpha}=0.01\) \\
\(P(\mu)\) & fixed \(T\) & \(T=50~{\rm MeV},\;\bar{\alpha}=0.01\) \\
\(\chi_{T,l}(T), \chi_{T,s}(T)\) & \(\mu=0\) & \(\bar{\alpha}=0.01\) \\
Light-sector boundary & \(T\)-\(\mu\) scan & crossover: \(\max\chi_{T,l}\);
first order: equal minima; CEP: meeting point\\
\hline
\end{tabular}
\end{table*}
At each temperature and chemical potential, the coupled light- and
strange-sector gap equations are solved iteratively until both
constituent masses satisfy the specified convergence tolerance. The
undeformed and GUP-deformed results are obtained with the same vacuum
parameter set listed in Table~\ref{tab:njl_parameters}, so that the
effect of the GUP sector arises solely from the modified
momentum-space measure, while the NJL couplings remain fixed.
Table~\ref{tab:numerical_protocol} likewise shows that all shifts
displayed in the following figures are produced by the deformation
parameter \(\alpha\), with all other numerical parameters held
unchanged.

The numerical values quoted in Table~\ref{tab:vacuum-check} can be verified
directly by substituting the condensates into the coupled mass equations. In
this comparison all dimensionful quantities are expressed in GeV, with
\(
G=\frac{1.835}{\Lambda^2}=5.05837~{\rm GeV}^{-2},
\,\,
K=\frac{12.36}{\Lambda^5}=155.939~{\rm GeV}^{-5},
\,\,
\Lambda=0.6023~{\rm GeV}.
\)
For the undeformed vacuum solution, using
\(\phi_l=-1.41630\times10^{-2}~{\rm GeV}^3\) and
\(\phi_s=-1.71112\times10^{-2}~{\rm GeV}^3\), Eq.~\eqref{eq:Ml_gup} gives
\begin{align}
M_l
&=m_l-4G\phi_l+2K\phi_l\phi_s \nonumber\\
&=5.500+286.567+75.582
=367.649~{\rm MeV}.
\end{align}
Similarly, Eq.~\eqref{eq:Ms_gup} gives
\begin{align}
M_s
&=m_s-4G\phi_s+2K\phi_l^2 \nonumber\\
&=140.700+346.219+62.560
=549.479~{\rm MeV}.
\end{align}
For the reference deformed case \(\bar{\alpha}=0.01\), using
\(\phi_l=-1.35411\times10^{-2}~{\rm GeV}^3\) and
\(\phi_s=-1.66422\times10^{-2}~{\rm GeV}^3\), the same substitution yields
\begin{align}
M_l
&=m_l-4G\phi_l+2K\phi_l\phi_s \nonumber\\
&=5.500+273.983+70.283
=349.766~{\rm MeV},
\label{eq.44}
\end{align}
and
\begin{align}
M_s
&=m_s-4G\phi_s+2K\phi_l^2 \nonumber\\
&=140.700+336.729+57.186
=534.616~{\rm MeV}.
\label{eq.45}
\end{align}
These direct substitutions show that the quoted vacuum masses and condensates
satisfy the coupled light-strange gap equations with the stated conventions.
They also clarify that the undeformed solution used in the numerical analysis
is \(M_l\simeq 367.65~{\rm MeV}\) and \(M_s\simeq 549.48~{\rm MeV}\), rather
than the rounded illustrative values used in the earlier table.

\begin{table}[htb]
\centering
\caption{\justifying Vacuum self-consistency comparison for the coupled
light-strange gap equations at \(T=\mu=0\). Condensates are quoted in
\({\rm GeV}^3\), while masses are quoted in MeV. The undeformed solution is
obtained with the Rehberg-Klevansky-H\"ufner parameters and serves as the
reference point for the deformed calculation.}
\label{tab:vacuum-check}
\begin{tabular}{c|cc}
\hline
Quantity & \(\bar{\alpha}=0\) & \(\bar{\alpha}=0.01\) \\
\hline
\(M_l\) & \(367.648\) & \(349.766\) \\
\(M_s\) & \(549.479\) & \(534.616\) \\
\(\phi_l\) & \(-1.41630\times10^{-2}\) & \(-1.35411\times10^{-2}\) \\
\(\phi_s\) & \(-1.71112\times10^{-2}\) & \(-1.66422\times10^{-2}\) \\
\(-\phi_l^{1/3}\) & \(241.946~{\rm MeV}\) & \(238.352~{\rm MeV}\) \\
\(-\phi_s^{1/3}\) & \(257.688~{\rm MeV}\) & \(255.311~{\rm MeV}\) \\
\hline
\end{tabular}
\end{table}

Since the deformation modifies the vacuum condensates, a fully
vacuum-constrained effective-model analysis would determine a clearly
specified subset of the vacuum parameters
\((m_l,m_s,G,K,\Lambda)\) independently at each nonzero \(\alpha\) by
requiring agreement with a chosen set of vacuum pseudoscalar
observables. In the present work, no such parameter determination is
undertaken, because the aim is to isolate the direct effect of the
deformed density of states while retaining the standard NJL parameter
set. The results should therefore be interpreted as a fixed-parameter
sensitivity analysis rather than as a precision description of vacuum
phenomenology at nonzero \(\alpha\). To quantify the deformation-induced
change in the vacuum sector, we use the leading
Gell-Mann-Oakes-Renner (GMOR) relation, \(m_\pi^2 f_\pi^2 \simeq -2m_l\phi_l\) \cite{Gell-Mann:1968hlm}.
For the vacuum diagnostic used here, \(f_\pi\) is evaluated at
\(T=\mu=0\) using the same fixed cutoff and phase-space factor as in the
gap equations 
\begin{equation}
f_\pi^2(\alpha)
=
\frac{N_c M_l^2(\alpha)}{2\pi^2}
\int_0^\Lambda dp\,
\frac{p^2 J_\alpha(p)}
{\left[p^2+M_l^2(\alpha)\right]^{3/2}},
\label{eq:fpi-vacuum}
\end{equation}
where \(M_l(\alpha)\) is the corresponding self-consistent vacuum
light-quark constituent mass, \(J_\alpha(p)=(1+\alpha p^2)^{-3}\), and
\(J_0(p)=1\). This gives
\(f_\pi^{\mathrm{NJL}}=92.391~\mathrm{MeV}\) and
\(f_\pi^{\mathrm{NJL+GUP}}=90.557~\mathrm{MeV}\) for
\(\bar{\alpha}=0.01\). Substitution of these values and the corresponding
vacuum condensates into the leading GMOR relation gives
\(m_\pi^{\mathrm{NJL}}=135.10~\mathrm{MeV}\) and
\(m_\pi^{\mathrm{NJL+GUP}}=134.77~\mathrm{MeV}\). Thus, the pion mass remains nearly unchanged at this deformation
strength, whereas the constituent masses and condensates exhibit the
expected changes induced by ultraviolet phase-space suppression. A future vacuum-constrained NJL+GUP analysis would be required before
using the model for precision vacuum phenomenology.

The equilibrium state is determined by the stationarity conditions in
Eq.~\eqref{eq:stationarity_gup_2p1}, which are equivalent to the coupled
deformed condensate equations given in Eqs.~\eqref{eq:phil_gup_rad} and
\eqref{eq:phis_gup_rad}. The constituent masses are then obtained
self-consistently through Eqs.~\eqref{eq:Ml_gup} and \eqref{eq:Ms_gup}.
Eqs.~\eqref{eq:phil_gup_rad}-\eqref{eq:Ms_gup} replace the single-gap equation
of the two-flavor formulation and form the main self-consistency system of the
present work. To obtain the leading small-\(\alpha\) behavior, we expand the
deformation factor for \(\alpha p^2\ll1\), as given in
Eq.~\eqref{eq:jacobian_expand_again_2p1}. Substituting this expansion
into Eqs.~\eqref{eq:phil_gup_rad} and
\eqref{eq:phis_gup_rad}, one obtains
\begin{equation}
\phi_l^{(\alpha)}(M_l)
=
\phi_{l,0}(M_l)
+
3\alpha\,\Delta_l(M_l)
+
O(\alpha^2),
\label{eq:phil_expand}
\end{equation}
\begin{equation}
\phi_s^{(\alpha)}(M_s)
=
\phi_{s,0}(M_s)
+
3\alpha\,\Delta_s(M_s)
+
O(\alpha^2),
\label{eq:phis_expand}
\end{equation}
where $\phi_{f,0}(M_f)$ denotes the undeformed condensate
functional evaluated at the same trial quasiparticle mass
$M_f$, with $f\in\{l,s\}$. Eqs.~\eqref{eq:phil_expand} and \eqref{eq:phis_expand} therefore
describe the explicit response of the momentum-space kernel
before the deformation-induced stationary shifts of $M_l$
and $M_s$ are included. The functions $\Delta_l(M_l)$ and
$\Delta_s(M_s)$ are
\begin{equation}
\Delta_l(M_l)
=
\frac{N_c}{\pi^2}
\int_0^\Lambda dp\,
p^4
\frac{M_l}{E_l}
\Big[
1-f_l^-(E_l)-f_l^+(E_l)
\Big],
\label{eq:Deltal_def}
\end{equation}
\begin{equation}
\Delta_s(M_s)
=
\frac{N_c}{\pi^2}
\int_0^\Lambda dp\,
p^4
\frac{M_s}{E_s}
\Big[
1-f_s^-(E_s)-f_s^+(E_s)
\Big].
\label{eq:Deltas_def}
\end{equation}
The signs of these corrections follow from the positivity of the thermal
kernel. For \(E_f\ge0\),
\begin{equation}
1-f_f^-(E_f)-f_f^+(E_f)\ge0,
\qquad
f\in\{l,s\},
\label{eq:kernel_positive_2p1}
\end{equation}
and therefore
\begin{equation}
\Delta_l>0,
\quad
\Delta_s>0,
\label{eq:Deltas_positive}
\end{equation}
on the broken branch. Since the physical condensates satisfy
\(\phi_{l,0}<0\) and \(\phi_{s,0}<0\), Eqs.~\eqref{eq:phil_expand} and
\eqref{eq:phis_expand} show that the GUP deformation reduces the magnitudes of
both condensates:
\begin{equation}
|\phi_l^{(\alpha)}|<|\phi_{l,0}|,
\qquad
|\phi_s^{(\alpha)}|<|\phi_{s,0}|,
\qquad
(\alpha>0).
\label{eq:phi_reduction}
\end{equation}
This result describes only the explicit change induced by the deformed
phase-space factor when the quasiparticle masses are held fixed. The complete
self-consistent solution also includes the induced variations of
\(M_l,\,M_s,\,\phi_l\), and \(\phi_s\). Consequently, the suppression of
high-momentum contributions by the deformed phase-space measure weakens
dynamical chiral symmetry breaking in both flavor sectors.

The same perturbative expansion can be applied to the fixed-cutoff
thermodynamic potential in
Eq.~\eqref{eq:gup_potential_radial}. At fixed condensates, one obtains Eq.~\eqref{eq:omega_small_alpha_2p1}. For
fixed \(\phi_l\) and \(\phi_s\), the leading correction is positive; hence the
deformation raises the grand potential and lowers the corresponding pressure.
Once the coupled stationary shifts in the condensates are included, the full
quantitative change must be obtained numerically, but the sign trend follows
from the expansion. 
The distinction between the explicit deformation of the
phase-space kernel and the induced stationary response can be
made exact at first order in $\alpha$. Away from a critical
singularity, the self-consistent masses and condensates are
expanded as
\begin{equation}
\begin{aligned}
M_f(\alpha)
=&
M_{f,0}
+
\alpha\,\delta M_f
+
O(\alpha^2),
\\
\phi_f^{(\alpha)}\!\left(M_f(\alpha)\right)
&=
\phi_{f,0}
+
\alpha\,\delta\phi_f
+
O(\alpha^2),
\end{aligned}
\label{eq.54}
\end{equation}
where \(f\in\{l,s\},\) and
\begin{equation}
\delta M_f
\equiv
\left.
\frac{\partial M_f}{\partial\alpha}
\right|_{\alpha=0},
\qquad
\delta\phi_f
\equiv
\left.
\frac{\partial\phi_f^{(\alpha)}}{\partial\alpha}
\right|_{\alpha=0}.
\label{eq.55}
\end{equation}
%Taylor expansion of Eqs.~\eqref{eq.44} and \eqref{eq.45}, including both the explicit measure dependence and the implicit dependence on the stationary quasiparticle masses, gives
Substituting the stationary-mass expansion in Eq.~\eqref{eq.54} into
the fixed-mass condensate expansions in Eqs.~\eqref{eq:phil_expand} and \eqref{eq:phis_expand},
and expanding the undeformed condensate functionals about
the stationary masses \(M_{f,0}\), gives, for
\(f\in\{l,s\}\),
\begin{equation}
%\[
\begin{aligned}
\phi_f^{(\alpha)}\!\left(M_f(\alpha)\right)
&=
\phi_{f,0}\!\left(M_{f,0}+\alpha\,\delta M_f\right)
+\\
&\hspace{2.0cm}3\alpha\,
\Delta_f\!\left(M_{f,0}+\alpha\,\delta M_f\right)
+\mathcal{O}(\alpha^2)
\\
&=
\phi_{f,0}(M_{f,0})
+\alpha
\Bigg[
\left.
\frac{\partial\phi_{f,0}(M_f)}
{\partial M_f}
\right|_{M_f=M_{f,0}}
\\
&\hspace{2.2cm} \delta M_f
+3\Delta_f(M_{f,0})
\Bigg]
+\mathcal{O}(\alpha^2).
\end{aligned}
%\]
\end{equation}
The dependence of \(\Delta_f\) on \(M_f(\alpha)\) contributes
only at order \(\alpha^2\), because \(\Delta_f\) is already
multiplied by the explicit factor \(\alpha\). Comparing this
expression with the definition of \(\delta\phi_f\) in
Eq.~\eqref{eq.55}, and using the response coefficient \(A_f\) defined
in Eq.~\eqref{eq.58}, one obtains
\begin{equation}
\delta\phi_l
=
3\Delta_l
+
\mathcal A_l\,\delta M_l,
\qquad
\delta\phi_s
=
3\Delta_s
+
\mathcal A_s\,\delta M_s,
\label{eq.56}
\end{equation}
where all quantities on the right-hand side are evaluated on
the undeformed stationary solution and
\begin{equation}
\mathcal A_f
\equiv
\left.
\frac{\partial\phi_{f,0}(M_f)}
     {\partial M_f}
\right|_{M_f=M_{f,0}}.
\label{eq.58}
\end{equation}
Hence we write
\begin{equation}
\mathcal K_f(E_f)
\equiv
1-f_f^-(E_f)-f_f^+(E_f),
\end{equation}
the response coefficient can be expressed explicitly, for
$T>0$, as
\begin{multline}
\mathcal A_f
=
-\frac{N_c}{\pi^2}
\int_0^\Lambda dp\,p^2
\Big[
\frac{p^2}{E_{f,0}^3}\,
\mathcal K_f(E_{f,0})
+\\
\frac{M_{f,0}^2}{T E_{f,0}^2}
\left\{
f_{f,0}^-\!\left(1-f_{f,0}^-\right)
+
f_{f,0}^+\!\left(1-f_{f,0}^+\right)
\right\}
\Big].
\label{eq.59}
\end{multline}
The $T\rightarrow0$ expression is understood as the continuous
zero-temperature limit. Since the integrand in square brackets
is non-negative, one has $\mathcal A_f<0$ on the physical
branch. 

Linearizing the coupled constituent-mass relations,
Eqs.~\eqref{eq:Ml_gup} and \eqref{eq:Ms_gup}, gives
\begin{equation}
\begin{pmatrix}
\delta M_l\\[1mm]
\delta M_s
\end{pmatrix}
=
\underbrace{
\begin{pmatrix}
-4G+2K\phi_{s,0} & 2K\phi_{l,0}\\
4K\phi_{l,0} & -4G
\end{pmatrix}}_{\displaystyle \mathcal B
\equiv
\begin{pmatrix}a&b\\c&d\end{pmatrix}}
\begin{pmatrix}
\delta\phi_l\\[1mm]
\delta\phi_s
\end{pmatrix}.
\label{eq.60}
\end{equation}
%Combining Eqs.~\eqref{eq:phi_reduction} and \eqref{eq.54}, the complete first-order self-consistent response obeys
Substituting the condensate variations in Eq.~\eqref{eq.56} into the linearized constituent-mass relations in Eq.~\eqref{eq.60}, the complete first-order self-consistent response obeys
\begin{equation}
\begin{aligned}
&
\begin{pmatrix}
1-a\mathcal A_l & -b\mathcal A_s\\
-c\mathcal A_l & 1-d\mathcal A_s
\end{pmatrix}
\begin{pmatrix}
\delta M_l\\
\delta M_s
\end{pmatrix}
=
\begin{pmatrix}
q_l\\
q_s
\end{pmatrix},
\\[1mm]
q_l
&\equiv
(\delta M_l)_{\rm ker}
=
3\left(a\Delta_l+b\Delta_s\right)
\\
&=
-12G\Delta_l
+
6K\left(
\phi_{s,0}\Delta_l+\phi_{l,0}\Delta_s
\right),
\\[1mm]
q_s
&\equiv
(\delta M_s)_{\rm ker}
=
3\left(c\Delta_l+d\Delta_s\right)
\\
&=
-12G\Delta_s
+
12K\phi_{l,0}\Delta_l .
\end{aligned}
\label{eq:kernel-source-system}
\end{equation}
Thus, the expressions previously written directly as
$\delta M_l$ and $\delta M_s$ are the explicit
phase-space kernel source terms $q_l$ and $q_s$, not the full
stationary mass shifts. Defining
\begin{equation}
\mathcal D
\equiv
(1-a\mathcal A_l)(1-d\mathcal A_s)
-
bc\,\mathcal A_l\mathcal A_s ,
\end{equation}
the full coupled solution is
\begin{equation}
\begin{aligned}
\delta M_l
&=
\frac{
(1-d\mathcal A_s)q_l
+
b\mathcal A_s q_s
}{\mathcal D},
\\[1mm]
\delta M_s
&=
\frac{
c\mathcal A_l q_l
+
(1-a\mathcal A_l)q_s
}{\mathcal D}.
\end{aligned}
\label{eq:full-mass-response}
\end{equation}
On the broken branch,
$\Delta_l>0$, $\Delta_s>0$,
$\phi_{l,0}<0$, and $\phi_{s,0}<0$, so that the direct kernel
sources satisfy
$q_l<0$ and $q_s<0$. The full mass shifts are obtained only
after inversion of the coupled response matrix in Eq.~\eqref{eq:kernel-source-system}.
For each regular stable branch retained in the numerical
analysis, the response matrix has positive principal minors,
\(
1-a\mathcal A_l>0,
\,\,
1-d\mathcal A_s>0,
\,\,
\mathcal D>0,
\)
and negative off-diagonal elements. It is therefore a
nonsingular $M$-matrix, whose inverse is element-wise
non-negative. Consequently,
\(
\delta M_l<0,
\,\,
\delta M_s<0
\)
for the stable broken and crossover branches considered here.
At a critical singularity, where $\mathcal D\rightarrow0$, the
linear response becomes singular and the perturbative
expression is not used. The phase structure is instead
obtained from the full nonlinear stationary grand potential.

To assess the accuracy of the first-order expansion, we consider
\(T=\mu=0\) and \(\bar{\alpha}=0.01\).
Eqs.~\eqref{eq.59}-\eqref{eq:kernel-source-system} yield
\(
\alpha\,\delta M_l=-18.11~\mathrm{MeV},
\,\,
\alpha\,\delta M_s=-15.16~\mathrm{MeV},
\)
whereas the complete nonlinear gap-equation solutions give
\(-17.882~\mathrm{MeV}\) and \(-14.864~\mathrm{MeV}\),
respectively. The close agreement shows that the reference deformation
is sufficiently small for the first-order expansion to remain accurate.
All figures and tables, however, are based on the complete nonlinear
self-consistent solutions. 
We first show the temperature dependence of the light-quark constituent mass at
vanishing chemical potential. Fig.~\ref{fig:MlT} shows that the GUP-deformed
curve lies systematically below the undeformed NJL result throughout the
chirally broken phase. This provides the numerical counterpart of the
analytical result that the deformation weakens dynamical mass generation. Near
the restoration region, the separation between the curves decreases as both
solutions approach their restored-phase values.

To examine the thermal behavior of the heavier flavor sector, we next
consider the temperature dependence of the strange-quark constituent mass.
Fig.~\ref{fig:MsT} shows that the positive-\(\alpha\) deformation also
reduces \(M_s\), although the reduction is smaller than in the light sector
because of the larger strange-quark current mass. The comparison between
Figs.~\ref{fig:MlT} and \ref{fig:MsT} therefore shows that the deformation
weakens dynamical chiral symmetry breaking in both sectors, but with different
quantitative strength.
%The density dependence of the effective masses conveys the same message. Figure~\ref{fig:Mmu} compares the chemical-potential dependence of the light and strange constituent masses at fixed temperature in the undeformed and GUP-deformed models. In both sectors, increasing chemical potential reduces the constituent mass, while the GUP deformation accelerates this reduction. The stronger sensitivity of the light sector is consistent with the earlier onset of light-sector chiral restoration inferred from the analytical discussion.
The chemical-potential dependence of the effective masses is shown in
Fig.~\ref{fig:Mmu}. The same self-consistent coupled gap-equation procedure
used for Figs.~\ref{fig:MlT} and \ref{fig:MsT} is used here as well. Because Fig.~\ref{fig:Mmu} presents the chemical-potential dependence
at fixed \(T=50~\mathrm{MeV}\), its \(\mu=0\) values can be compared
directly with the corresponding \(T=50~\mathrm{MeV}\) values shown in
Figs.~\ref{fig:MlT} and \ref{fig:MsT}. In the undeformed model one obtains
\(M_l=367.571~\mathrm{MeV}\) and \(M_s=549.444~\mathrm{MeV}\) at
\((T,\mu)=(50~{\rm MeV},0)\), while for \(\bar{\alpha}=0.01\) one obtains
\(M_l=349.661~\mathrm{MeV}\) and \(M_s=534.568~\mathrm{MeV}\). These values
coincide with the corresponding \(T=50~\mathrm{MeV}\) points of
Figs.~\ref{fig:MlT} and \ref{fig:MsT}, as summarized in
Table~\ref{tab:numerical_comparison_corrected}. With increasing chemical
potential, the light-sector mass remains close to its broken-phase value at
small \(\mu\), and then decreases rapidly in the low-temperature restoration
region. The strange-sector mass changes through the coupled light-strange gap
equations. Since the calculation is performed at \(T=50~\mathrm{MeV}\), the
low-temperature solution may exhibit rapid branch changes rather than a purely
smooth crossover. The sharp light-sector drop is transmitted to the strange
sector through the Kobayashi-Maskawa-'t~Hooft determinant coupling, while the
additional rapid decrease of \(M_s\) at larger \(\mu\) reflects the approach to
the dense strange branch. The symbols denote the self-consistent solutions at
the displayed chemical potentials, and the lines guide the eye through the
resolved branches. The GUP-deformed curves lie below the undeformed ones,
consistent with the reduction of the scalar condensates and the weakening of
dynamical chiral symmetry breaking induced by the ultraviolet-reduced
phase-space measure.
The deformation-induced reduction of the light constituent mass is
reflected directly in the chiral phase structure. At sufficiently high
temperature, the light-sector transition is a smooth crossover. In this
region, the pseudocritical temperature is defined by the maximum of
\(\chi_{T,l}=-\partial M_l/\partial T\), equivalently by the inflection
point of \(M_l(T,\mu)\). At lower temperature, however, the stationary
grand potential develops competing chirally broken and approximately
restored local minima, and the physical transition becomes first order.
The first-order coexistence line is therefore determined from the
degeneracy condition
\begin{equation}
\Omega^\star_{\rm broken}(T,\mu)
=
\Omega^\star_{\rm restored}(T,\mu),
\label{eq:coexistence}
\end{equation}
rather than from an inflection-point criterion. The crossover and
first-order branches meet at the light-sector critical endpoint. 
Within the fixed-parameter calculation, the deformation-induced movement
of the two portions of the chiral boundary may be summarized as
\begin{equation}
\begin{aligned}
T_{\rm pc}^{\rm NJL+GUP}(\mu)
&<
T_{\rm pc}^{\rm NJL}(\mu),
&&\text{on the crossover branch},\\
\mu_{\chi}^{\rm NJL+GUP}(T)
&<
\mu_{\chi}^{\rm NJL}(T),
&&\text{on the first-order branch}.
\end{aligned}
\label{eq:phase-shifts}
\end{equation} 
Fig.~\ref{fig:phase} shows the resulting light-sector chiral
transition boundary. The dashed lines denote the crossover branches,
located from the peak of \(\chi_{T,l}\), whereas the solid lines denote
the first-order coexistence branches, obtained from
Eq.~\eqref{eq:coexistence}. For each theory, the crossover and
first-order branches meet at the corresponding critical endpoint. In
the undeformed NJL model, the critical endpoint is located at
\(
(\mu_{\rm CEP},T_{\rm CEP})_{\rm NJL}
=
(318.43,67.72)\ {\rm MeV},
\)
whereas in the GUP-deformed model with
\(\bar{\alpha}=0.01\) it is located at
\(
(\mu_{\rm CEP},T_{\rm CEP})_{\rm NJL+GUP}
=
(313.97,60.20)\ {\rm MeV}.
\)
Thus, within the present fixed-parameter analysis, the positive-\(\alpha\)
deformation shifts the crossover segment, the first-order coexistence
segment, and the critical endpoint toward lower temperature and lower
chemical potential. 
As an explicit consistency check with Fig.~\ref{fig:Mmu}, the
first-order coexistence chemical potentials at
\(T=50~\mathrm{MeV}\) are
\(
\mu_{\chi}^{\rm NJL}
=
335.26~\mathrm{MeV},
\,\,
\mu_{\chi}^{\rm NJL+GUP}
=
323.22~\mathrm{MeV}.
\)
These values coincide with the discontinuous light-sector branch changes
shown in the fixed-temperature mass solutions. Consequently, the
\(T=50~\mathrm{MeV}\) points in Fig.~\ref{fig:phase} belong to the
first-order coexistence branches and are not identified through a
crossover inflection-point condition.

\begin{figure}[t]
\centering
\begin{tikzpicture}
\begin{axis}[
    width=\linewidth,
    height=7.2cm,
    xlabel={$\mu\,[\mathrm{MeV}]$},
    ylabel={$T_{\chi}\,[\mathrm{MeV}]$},
    xmin=0,
    xmax=350,
    ymin=40,
    ymax=210,
    xtick={0,50,100,150,200,250,300,350},
    ytick={50,75,100,125,150,175,200},
    minor tick num=1,
    legend style={
        draw=none,
        fill=none,
        at={(0.58,0.38)},
        anchor=north east,
        font=\scriptsize,
        legend columns=1,
        legend cell align=left,
        /tikz/every even column/.append style={column sep=0.7em}
    },
    tick label style={font=},%\small},
    label style={font=},%\small},
    axis line style={black},
    tick style={black},
    line join=round,
    line cap=round,
    clip marker paths=true
]

%------------------------------------------------
% NJL crossover branch:
% obtained from the peak of chi_{T,l}
%------------------------------------------------
\addplot[
    very thick,
    dashed,
    color=njlblue,
    mark=*,
    mark size=1.6pt,
    mark options={
        solid,
        fill=njlblue,
        draw=njlblue
    }
] table[row sep=\\] {
mu       Tc \\
0        195.90 \\
100      186.20 \\
200      152.80 \\
250      123.00 \\
300       84.80 \\
318.43    67.72 \\
};
\addlegendentry{NJL: crossover}

%------------------------------------------------
% NJL first-order coexistence branch:
% obtained from equality of stationary minima
%------------------------------------------------
\addplot[
    very thick,
    solid,
    color=njlblue,
    mark=triangle*,
    mark size=1.8pt,
    mark options={
        solid,
        fill=njlblue,
        draw=njlblue
    }
] table[row sep=\\] {
mu       Tc \\
318.43   67.72 \\
321.11   65.00 \\
325.96   60.00 \\
330.70   55.00 \\
335.26   50.00 \\
};
\addlegendentry{NJL: first order}

%------------------------------------------------
% NJL+GUP crossover branch:
% obtained from the peak of chi_{T,l}
%------------------------------------------------
\addplot[
    very thick,
    dashed,
    color=gupred,
    mark=square*,
    mark size=1.7pt,
    mark options={
        solid,
        fill=gupred,
        draw=gupred
    }
] table[row sep=\\] {
mu       Tc \\
0        188.70 \\
100      178.80 \\
200      144.20 \\
250      113.40 \\
300       73.80 \\
313.97    60.20 \\
};
\addlegendentry{NJL+GUP: crossover}

%------------------------------------------------
% NJL+GUP first-order coexistence branch:
% obtained from equality of stationary minima
%------------------------------------------------
\addplot[
    very thick,
    solid,
    color=gupred,
    mark=diamond*,
    mark size=1.9pt,
    mark options={
        solid,
        fill=gupred,
        draw=gupred
    }
] table[row sep=\\] {
mu       Tc \\
313.97   60.20 \\
316.02   58.00 \\
318.77   55.00 \\
321.46   52.00 \\
323.22   50.00 \\
};
\addlegendentry{NJL+GUP: first order}

%------------------------------------------------
% Critical endpoints
%------------------------------------------------
\addplot[
    only marks,
    color=njlblue,
    mark=star,
    mark size=3.1pt,
    mark options={
        solid,
        fill=njlblue,
        draw=njlblue
    },
    forget plot
] coordinates {
    (318.43,67.72)
};

\addplot[
    only marks,
    color=gupred,
    mark=star,
    mark size=3.1pt,
    mark options={
        solid,
        fill=gupred,
        draw=gupred
    },
    forget plot
] coordinates {
    (313.97,60.20)
};

\end{axis}
\end{tikzpicture}

\caption{\justifying
Light-sector chiral transition boundary in the $T$--$\mu$ plane for
the undeformed NJL model and the GUP-deformed model with
$\bar{\alpha}=0.01$. The dashed segments denote the crossover
branches, whose pseudocritical temperatures are determined from the
maximum of
$\chi_{T,l}=-\partial M_l/\partial T$, equivalently from the inflection
point of $M_l(T,\mu)$. The solid segments denote the first-order
coexistence branches, obtained by requiring equality of the competing
broken- and restored-branch stationary grand potentials, Eq.~\eqref{eq:coexistence}.
The stars identify the critical endpoints:
$(\mu_{\rm CEP},T_{\rm CEP})
 =(318.43,67.72)~\mathrm{MeV}$ for the undeformed NJL model and
$(313.97,60.20)~\mathrm{MeV}$ for the GUP-deformed model.
At $T=50~\mathrm{MeV}$, the first-order transition chemical potentials
are
$\mu_{\chi}^{\rm NJL}=335.26~\mathrm{MeV}$ and
$\mu_{\chi}^{\rm NJL+GUP}=323.22~\mathrm{MeV}$, consistently reproducing
the discontinuous branch changes shown in Fig.~\ref{fig:Mmu}.
The positive-$\alpha$ deformation shifts the crossover branch, the
first-order coexistence branch, and the critical endpoint toward lower
temperature and lower chemical potential within the fixed-parameter
analysis.}
\label{fig:phase}
\end{figure}
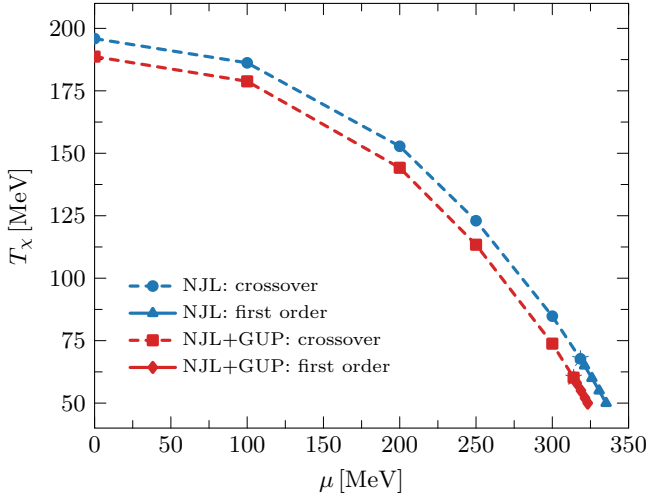
Once the self-consistent stationary solution
\((\phi_l^\star,\phi_s^\star)\) is known, the thermodynamic quantities follow
from the stationary grand potential. We distinguish two pressure definitions:
\begin{equation}
P^\star(T,\mu,\alpha)\equiv-\Omega^\star(T,\mu,\alpha),
\end{equation}
and
\begin{equation}
P_{\rm sub}(T,\mu,\alpha)\equiv-
\left[
\Omega^\star(T,\mu,\alpha)
-
\Omega^\star(0,0,\alpha)
\right].
\end{equation} 
The small-\(\alpha\) expansion at fixed condensates following
Eq.~\eqref{eq:omega_small_alpha_2p1} refers to \(P^\star\), the unsubtracted
stationary pressure. The pressure shown in Figs.~\ref{fig:PT} and
\ref{fig:Pmu} is instead \(P_{\rm sub}\), where the vacuum contribution is
subtracted separately in each theory. Since the positive-\(\alpha\)
deformation also changes the vacuum value of the stationary grand potential, a
small increase of \(P_{\rm sub}\) is compatible with a reduction of
\(P^\star\). 
The quark number density, entropy density, and energy density are obtained from
\begin{equation}
n_q(T,\mu,\alpha)
=
-
\frac{\partial\Omega^\star}{\partial\mu},
\quad
s(T,\mu,\alpha)
=
-
\frac{\partial\Omega^\star}{\partial T},
\end{equation}
and the vacuum-subtracted energy density is
\begin{equation}
\epsilon(T,\mu,\alpha)
=
-
P_{\rm sub}
+
Ts
+
\mu n_q.
\end{equation}
Fig.~\ref{fig:PT} shows the temperature dependence of \(P_{\rm sub}\) at
\(\mu=0\). At \((T,\mu)=(200~\mathrm{MeV},0)\), the values are
\begin{equation}
P_{\rm sub}^{\rm NJL}
=
0.210~\mathrm{GeV/fm}^3,
\,\,
P_{\rm sub}^{\rm NJL+GUP}
=
0.220~\mathrm{GeV/fm}^3.
\end{equation}
At the same point, the unsubtracted stationary pressures are
\begin{equation}
P^{\star}_{\rm NJL}
=
4.474~\mathrm{GeV/fm}^3,
\,\,
P^{\star}_{\rm NJL+GUP}
=
4.383~\mathrm{GeV/fm}^3.
\end{equation}
Thus the positive-\(\alpha\) deformation reduces the unsubtracted stationary
pressure, while the vacuum-subtracted pressure is slightly larger because the
vacuum contribution is also changed by the deformed phase-space measure.
Fig.~\ref{fig:Pmu} shows the corresponding chemical-potential dependence of
\(P_{\rm sub}\) at \(T=50~\mathrm{MeV}\). With the same vacuum-subtraction
convention, the GUP-deformed \(P_{\rm sub}\) is slightly larger than the
undeformed value over the range shown. This follows from subtracting the vacuum
contribution separately in each theory and should not be confused with the
fixed-condensate ultraviolet reduction of \(P^\star\).
\begin{figure}[htb]
\centering
\begin{tikzpicture}
\begin{axis}[
    width=\linewidth,
    height=7.2cm,
    xlabel={$T\,[\mathrm{MeV}]$},
    ylabel={$P_{\rm sub}\,[\mathrm{GeV/fm^3}]$},
    xmin=0,
    xmax=250,
    ymin=0,
    ymax=0.52,
    xtick={0,50,100,150,200,250},
    ytick={0,0.1,0.2,0.3,0.4,0.5},
    minor tick num=1,
    legend style={
        draw=none,
        fill=none,
        at={(0.03,0.97)},
        anchor=north west,
        font=\small,
        legend cell align=left
    },
    tick label style={font=},%\small},
    label style={font=},%\small},
    axis line style={black},
    tick style={black},
    line join=round,
    line cap=round,
    clip marker paths=true
]

\addplot[
    very thick,
    dashed,
    color=njlblue,
    mark=*,
    mark size=1.6pt,
    mark options={solid, fill=njlblue, draw=njlblue}
] table[row sep=\\] {
T     P \\
0     0.000000 \\
100   0.005451 \\
150   0.054333 \\
175   0.114920 \\
200   0.209693 \\
225   0.336749 \\
250   0.486593 \\
};
\addlegendentry{NJL}

\addplot[
    very thick,
    color=gupred,
    mark=square*,
    mark size=1.8pt,
    mark options={solid, fill=gupred, draw=gupred}
] table[row sep=\\] {
T     P \\
0     0.000000 \\
100   0.006167 \\
150   0.058718 \\
175   0.122634 \\
200   0.220476 \\
225   0.347273 \\
250   0.495464 \\
};
\addlegendentry{NJL+GUP}

\end{axis}
\end{tikzpicture}
\caption{\justifying
Temperature dependence of the vacuum-subtracted pressure
$P_{\rm sub}$ at $\mu=0$ in the undeformed NJL model and in the
GUP-deformed model with $\bar{\alpha}=0.01$. Each plotted point is
obtained from the same self-consistent stationary solution of the coupled
gap equations used throughout the revised analysis, with the vacuum
contribution subtracted separately in each theory, Eq.~\eqref{eq.40}. 
%$P_{\rm sub}(T,\mu)\equiv-[\Omega^\star(T,\mu)-\Omega^\star(0,0)]$.
At $T=200~\mathrm{MeV}$, the values are
$P_{\rm sub}^{\rm NJL}=0.209693~\mathrm{GeV/fm^3}$ and
$P_{\rm sub}^{\rm NJL+GUP}=0.220476~\mathrm{GeV/fm^3}$.
At $T=250~\mathrm{MeV}$, they are
$0.486593~\mathrm{GeV/fm^3}$ and
$0.495464~\mathrm{GeV/fm^3}$, respectively.
This figure displays the vacuum-subtracted pressure and therefore should
be distinguished from the unsubtracted stationary pressure
$P^\star\equiv-\Omega^\star$. The small increase of $P_{\rm sub}$ in the
deformed theory is a consequence of the deformation-induced shift of the
vacuum baseline and does not contradict the fixed-condensate ultraviolet
reduction of the unsubtracted stationary pressure.}
\label{fig:PT}
\end{figure}
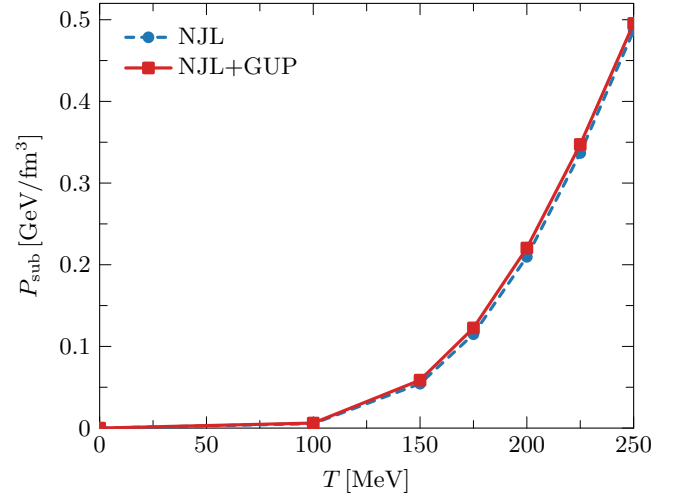
\begin{figure}[htb]
\centering
\begin{tikzpicture}
\begin{axis}[
    width=\linewidth,
    height=7.2cm,
    xlabel={$\mu\,[\mathrm{MeV}]$},
    ylabel={$P_{\rm sub}\,[\mathrm{GeV/fm^3}]$},
    xmin=200, xmax=500,
    ymin=0, ymax=0.45,
    legend style={
        draw=none,
        fill=none,
        at={(0.03,0.97)},
        anchor=north west,
        font=\small
    },
    tick label style={font=},%\small},
    label style={font=},%\small},
    axis line style={black},
    tick style={black},
    line join=round,
    line cap=round
]

\addplot[
    very thick,
    dashed,
    color=njlblue,
    mark=*,
    mark size=1.4pt,
    mark options={fill=njlblue}
] table[row sep=\\] {
mu   P \\
0    0.000020 \\
100  0.000077 \\
200  0.000564 \\
300  0.004552 \\
330  0.009392 \\
340  0.016437 \\
350  0.029135 \\
400  0.111480 \\
450  0.230700 \\
500  0.411797 \\
};
\addlegendentry{\textcolor{njlblue}{NJL}}

\addplot[
    very thick,
    color=gupred,
    mark=square*,
    mark size=1.6pt,
    mark options={fill=gupred}
] table[row sep=\\] {
mu   P \\
0    0.000027 \\
100  0.000103 \\
200  0.000755 \\
300  0.006339 \\
330  0.018775 \\
340  0.030378 \\
350  0.043180 \\
400  0.125249 \\
450  0.244360 \\
500  0.426620 \\
};
\addlegendentry{\textcolor{gupred}{NJL+GUP}}

\end{axis}
\end{tikzpicture}
\caption{\justifying Chemical-potential dependence of the vacuum-subtracted pressure \(P_{\rm sub}\)
at fixed temperature \(T=50~\mathrm{MeV}\) in the undeformed NJL model and in
the GUP-deformed model with \(\bar{\alpha}=0.01\). The quantity
\(P_{\rm sub}\) is defined in Eq.~\eqref{eq.40}, with the vacuum contribution
subtracted separately in each theory. With this convention, the GUP-deformed
curve lies slightly above the undeformed curve. For example, at
\(\mu=500~\mathrm{MeV}\),
\(P_{\rm sub}^{\rm NJL}=0.411797~\mathrm{GeV/fm^3}\), while
\(P_{\rm sub}^{\rm NJL+GUP}=0.426620~\mathrm{GeV/fm^3}\). This behavior follows
from the change in the vacuum value of the stationary grand potential induced
by the deformed phase-space measure and is consistent with the reduction of the
unsubtracted stationary pressure \(P^\star=-\Omega^\star\).
}
\label{fig:Pmu}
\end{figure}
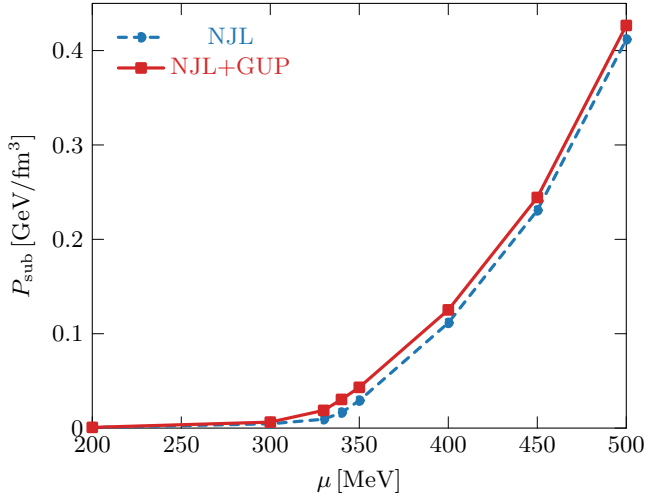
To examine the zero-temperature behavior under compact-star
equilibrium conditions, we now solve the complete flavor-resolved
mean-field problem. Although the current masses satisfy
$m_u=m_d\equiv m_l$, beta equilibrium produces an isospin-asymmetric
medium because $\mu_u\neq\mu_d$ whenever $\mu_e\neq0$. The light-sector
reduction $\phi_u=\phi_d\equiv\phi_l$ is therefore not imposed in this
subsection. Instead, the three scalar condensates
$\phi_u$, $\phi_d$, and $\phi_s$ are determined independently. 
At $T=0$, the flavor-resolved quark grand-potential density, evaluated
with the same common three-momentum cutoff and the same deformed
density of states factor used throughout the manuscript, is
\begin{multline}
\Omega_q
={}
2G\left(\phi_u^2+\phi_d^2+\phi_s^2\right)
-4K\phi_u\phi_d\phi_s
\\
-\frac{N_c}{\pi^2}
\sum_{f=u,d,s}
\int_0^\Lambda dp\,
p^2 J_\alpha(p)\times
\\
\left[
E_f+
\left(\mu_f-E_f\right)
\Theta\!\left(\mu_f-E_f\right)
\right],
\label{eq:beta_omega_q}
\end{multline}
where \(J_\alpha(p)\) is given in Eq.~\eqref{eq.gup_measure} of Sec.~\ref{sec:gup_measure} and \(E_f(p)=\sqrt{p^2+M_f^2}\) is the energy of different quasiparticles. 
The complete flavor-resolved constituent-mass relations are
\begin{equation}
\begin{aligned}
M_u &= m_u-4G\phi_u+2K\phi_d\phi_s,
\\
M_d &= m_d-4G\phi_d+2K\phi_u\phi_s,
\\
M_s &= m_s-4G\phi_s+2K\phi_u\phi_d.
\label{eq:beta_masses}
\end{aligned}
\end{equation}
The corresponding zero-temperature condensate equations are
\begin{equation}
\phi_f=
-\frac{N_c}{\pi^2}
\int_0^\Lambda dp\,
p^2 J_\alpha(p)
\frac{M_f}{E_f}
\left[
1-\Theta\!\left(\mu_f-E_f\right)
\right],
\label{eq:beta_condensates}
\end{equation}
where \(f=u,d,s.\) Eqs.~\eqref{eq:beta_masses} and
\eqref{eq:beta_condensates} retain the full
Kobayashi-Maskawa-'t Hooft flavor mixing while allowing the
$u$- and $d$-sector condensates and constituent masses to respond
independently to their different chemical potentials
\cite{Rehberg:1995kh,Blaschke:2005uj}. 
For cold neutrino-free matter, weak equilibrium requires
\begin{equation}
\mu_d=\mu_s=\mu_u+\mu_e.
\label{eq:beta_equilibrium}
\end{equation}
In terms of the baryon and electron chemical potentials,
\begin{equation}
\mu_u=\frac{\mu_B}{3}-\frac{2\mu_e}{3},
\qquad
\mu_d=\mu_s=\frac{\mu_B}{3}+\frac{\mu_e}{3}.
\label{eq:beta_chemical_potentials}
\end{equation}
Local electric-charge neutrality is imposed through
\begin{equation}
\frac{2}{3}n_u-\frac{1}{3}n_d-\frac{1}{3}n_s-n_e=0,
\label{eq:beta_neutrality}
\end{equation}
and the baryon density is
\begin{equation}
n_B=\frac{n_u+n_d+n_s}{3}.
\label{eq:beta_baryon_density}
\end{equation} 
At the stationary solution, the quark number densities follow from
the explicit chemical-potential derivatives of the same grand
potential,
\begin{equation}
n_f=
-\left.
\frac{\partial\Omega_q}{\partial\mu_f}
\right|_{\phi_u^\star,\phi_d^\star,\phi_s^\star}
=
\frac{N_c}{\pi^2}
\int_0^{p_{F,f}^{\Lambda}}
dp\,
p^2J_\alpha(p),
\label{eq:beta_densities}
\end{equation}
where
\begin{equation}
p_{F,f}^{\Lambda}=
\begin{cases}
\displaystyle
\min\left\{\sqrt{\mu_f^2-M_f^2},\,\Lambda\right\},
& \mu_f>M_f,\\[2mm]
0, & \mu_f\leq M_f.
\end{cases}
\label{eq:beta_fermi_momentum}
\end{equation}
Electrons are treated as a free relativistic gas,
\begin{equation}
\Omega_e=-\frac{\mu_e^4}{12\pi^2},
\qquad
n_e=\frac{\mu_e^3}{3\pi^2},
\qquad
P_e=\frac{\mu_e^4}{12\pi^2}.
\label{eq:beta_electrons}
\end{equation}
The electron mass may be neglected at the solutions quoted below
because $\mu_e\gg m_e$. 
The total stationary grand potential and vacuum-subtracted pressure
are
\begin{align}
\Omega_{\rm tot}^\star(\mu_B,\alpha)
&=
\Omega_q^\star(\mu_B,\alpha)+\Omega_e(\mu_e),
\label{eq:beta_omega_total}\\
P_{\rm tot}(\mu_B,\alpha)
&=
-\left[
\Omega_{\rm tot}^\star(\mu_B,\alpha)
-\Omega_{\rm tot}^\star(0,\alpha)
\right].
\label{eq:beta_pressure}
\end{align}
At each value of $\mu_B$, the four quantities
$\phi_u$, $\phi_d$, $\phi_s$, and $\mu_e$ are determined
simultaneously from the three condensate equations and the charge
neutrality condition. In regions with more than one stationary
solution, continuation from both the low-density and high-density
sides and multiple initial seeds are used, and the solution with the
lowest stationary grand potential is retained. 
The vacuum-subtracted energy density is
\begin{equation}
\epsilon_{\rm tot}
=
-P_{\rm tot}
+\sum_{f=u,d,s}\mu_fn_f+\mu_en_e.
\label{eq:beta_energy_density}
\end{equation}
Using Eqs.~\eqref{eq:beta_chemical_potentials} and
\eqref{eq:beta_neutrality}, one obtains the exact identity
\begin{equation}
\sum_{f=u,d,s}\mu_fn_f+\mu_en_e
=
\mu_B n_B.
\label{eq:beta_mu_identity}
\end{equation}
Consequently,
\begin{equation}
\epsilon_{\rm tot}=-P_{\rm tot}+\mu_B n_B,
\qquad
\frac{E}{A}
=
\frac{\epsilon_{\rm tot}}{n_B}
=
\mu_B-\frac{P_{\rm tot}}{n_B}.
\label{eq:beta_EA}
\end{equation}
At a finite-density zero-pressure point,
\begin{equation}
P_{\rm tot}=0
\quad\Longrightarrow\quad
\frac{E}{A}=\mu_B.
\label{eq:beta_zero_pressure}
\end{equation}
Moreover, along the stable charge-neutral beta-equilibrated sequence,
\begin{equation}
P_{\rm tot}
=
n_B^2
\frac{d}{dn_B}
\left(\frac{E}{A}\right),
\label{eq:beta_thermo_identity}
\end{equation}
so the minimum of $E/A$ coincides with the finite-density
zero-pressure point.

Solving the full three-condensate, charge-neutral, beta-equilibrated
system with the parameter set in Table~\ref{tab:njl_parameters} gives the zero-temperature
results summarized in Table~\ref{tab:beta_equilibrium}.
%Solving the full three-condensate system with the parameter set in Table~\ref{tab:njl_parameters} gives the results summarized in Table~\ref{tab:numerical_comparison_corrected}. 
The equality between
the zero-pressure density and the density at the minimum of $E/A$ is
satisfied within the displayed numerical precision. The conventional
absolute-stability requirement relative to iron is
\begin{equation}
\left(\frac{E}{A}\right)_{\min}<930~{\rm MeV}.
\label{eq:beta_absolute_stability}
\end{equation}
This condition is not satisfied in either theory. 
A further composition check is important. At the finite-density zero-pressure points one finds \(\mu_s < M_s\), so
that the strange-quark Fermi momentum vanishes,
\(p_{F,s}=0\), and consequently the strange-quark number density also
vanishes, \(n_s=0\). The minimum-energy states are consequently
charge-neutral beta-equilibrated quark-matter states with an
unpopulated strange Fermi sea, rather than strange-quark-matter
ground states. In the undeformed theory the strange Fermi sea begins
to populate at
$\mu_B\simeq1294.90~{\rm MeV}$ and
$n_B\simeq0.655~{\rm fm}^{-3}$, while for
$\bar{\alpha}=0.01$ the corresponding onset occurs at
$\mu_B\simeq1272.36~{\rm MeV}$ and
$n_B\simeq0.615~{\rm fm}^{-3}$. 
The positive-$\alpha$ phase-space factor establishes a suppression of
high-momentum state counting. Since
\begin{equation}
J_\alpha(p)=\frac{1}{(1+\alpha p^2)^3}<1,
\qquad \alpha>0,\quad p>0,
\label{eq:measure_suppression}
\end{equation}
high-momentum quark states carry less statistical weight than in the
undeformed NJL model. At fixed mean fields, this suppresses their
contribution to the unnormalised stationary pressure
$P^\star=-\Omega^\star$. It does not, however, establish by itself
that the equation of state is softer. Equation-of-state stiffness is
defined through a relation such as $P(\epsilon)$ or, locally, through
the squared speed of sound
\(
c_s^2=\frac{dP}{d\epsilon}.
\)
Moreover, Figs.~\eqref{fig:PT} and~\eqref{fig:Pmu} show that the vacuum-subtracted pressure
$P_{\rm sub}$ is slightly larger in the deformed theory because the
deformation also shifts the vacuum value of the stationary grand
potential. A quantitative comparison of equation of state stiffness
would therefore require a direct calculation of $P(\epsilon)$ or
$c_s^2$, which is beyond the scope of the present analysis. The
inclusion of a repulsive vector channel in a future vector-NJL or
vector-PNJL+GUP extension would permit the combined influence of
minimal-length phase-space suppression and vector repulsion on
$P(\epsilon)$ to be studied explicitly. A sign-reversed or pole-like
deformation is not considered because it would not represent the
positive-$\alpha$ minimal-length ansatz adopted here and could
introduce singular behavior within the NJL cutoff interval.

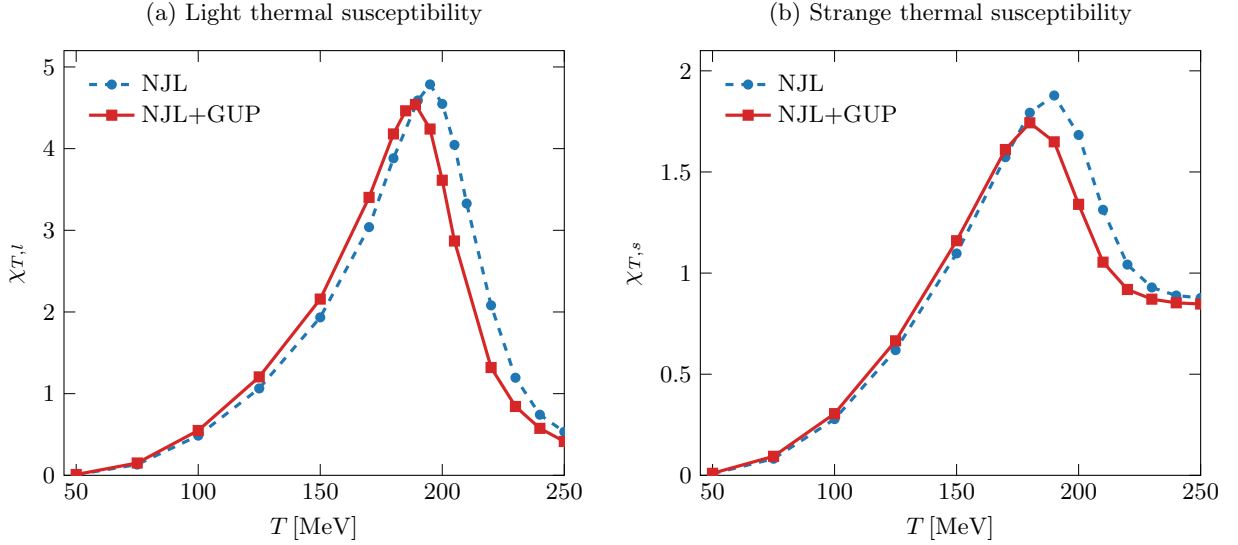
\begin{figure*}[htb]
\centering
\begin{tikzpicture}
\begin{groupplot}[
    group style={group size=2 by 1, horizontal sep=1.8cm},
    width=8.2cm,%\linewidth,
    height=7.2cm,
    xlabel={$T\,[\mathrm{MeV}]$},
    xtick={50,100,150,200,250},
    tick label style={font=},%\small},
    label style={font=},%\small},
    legend style={
        draw=none,
        fill=none,
        font=\small,
        legend cell align=left,
        at={(0.03,0.97)},
        anchor=north west
    },
    axis line style={black},
    tick style={black},
    clip marker paths=true
]

\nextgroupplot[
    ylabel={$\chi_{T,l}$},
    xmin=45, xmax=250,
    ymin=0, ymax=5.2,
    ytick={0,1,2,3,4,5},
    title={(a) Light thermal susceptibility}
]
\addplot[
    very thick,
    dashed,
    color=njlblue,
    mark=*,
    mark size=1.3pt,
    mark options={solid, fill=njlblue, draw=njlblue}
] table[row sep=\\] {
T    chi \\
50   0.008 \\
75   0.132 \\
100  0.486 \\
125  1.064 \\
150  1.934 \\
170  3.039 \\
180  3.882 \\
190  4.590 \\
195  4.786 \\
200  4.548 \\
205  4.046 \\
210  3.328 \\
220  2.081 \\
230  1.195 \\
240  0.742 \\
250  0.533 \\
};
\addlegendentry{NJL}

\addplot[
    very thick,
    color=gupred,
    mark=square*,
    mark size=1.5pt,
    mark options={solid, fill=gupred, draw=gupred}
] table[row sep=\\] {
T    chi \\
50   0.010 \\
75   0.151 \\
100  0.549 \\
125  1.205 \\
150  2.158 \\
170  3.403 \\
180  4.180 \\
185  4.463 \\
189  4.540 \\
195  4.240 \\
200  3.613 \\
205  2.869 \\
220  1.320 \\
230  0.843 \\
240  0.575 \\
250  0.415 \\
};
\addlegendentry{NJL+GUP}

\nextgroupplot[
    ylabel={$\chi_{T,s}$},
    xmin=45, xmax=250,
    ymin=0, ymax=2.1,
    ytick={0,0.5,1.0,1.5,2.0},
    title={(b) Strange thermal susceptibility}
]
\addplot[
    very thick,
    dashed,
    color=njlblue,
    mark=*,
    mark size=1.3pt,
    mark options={solid, fill=njlblue, draw=njlblue}
] table[row sep=\\] {
T    chi \\
50   0.008 \\
75   0.082 \\
100  0.278 \\
125  0.619 \\
150  1.097 \\
170  1.573 \\
180  1.793 \\
190  1.878 \\
200  1.683 \\
210  1.313 \\
220  1.042 \\
230  0.929 \\
240  0.889 \\
250  0.877 \\
};
\addlegendentry{NJL}

\addplot[
    very thick,
    color=gupred,
    mark=square*,
    mark size=1.5pt,
    mark options={solid, fill=gupred, draw=gupred}
] table[row sep=\\] {
T    chi \\
50   0.010 \\
75   0.094 \\
100  0.305 \\
125  0.665 \\
150  1.160 \\
170  1.611 \\
180  1.743 \\
190  1.649 \\
200  1.340 \\
210  1.054 \\
220  0.919 \\
230  0.871 \\
240  0.853 \\
250  0.847 \\
};
\addlegendentry{NJL+GUP}

\end{groupplot}
\end{tikzpicture}
\caption{\justifying
Temperature dependence of the light- and strange-sector thermal chiral-susceptibility
functions
$\chi_{T,l}=-\partial M_l/\partial T$ and
$\chi_{T,s}=-\partial M_s/\partial T$
at $\mu=0$ in the undeformed NJL model and in the controlled GUP-deformed
model with $\bar{\alpha}=0.01$.
The plotted points are obtained numerically from the corrected mass curves of
Figs.~\ref{fig:MlT} and \ref{fig:MsT} using finite-difference derivatives of the same self-consistent
stationary solutions of the coupled gap equations. In particular, the
high-temperature tail of the GUP light-sector response has been recomputed
directly from the corrected $M_l(T)$ solution, so that the plotted values are
fully consistent with the propagated mass curve. In the light sector, the
response peak shifts from
$T_{\mathrm{pc}}^{\mathrm{NJL}}\simeq 195.9~\mathrm{MeV}$ to
$T_{\mathrm{pc}}^{\mathrm{NJL+GUP}}\simeq 188.7~\mathrm{MeV}$, consistent
with the downward shift of the crossover temperature. The strange-sector
response is broader and weaker, reflecting the larger explicit strange-quark
mass and the smoother thermal evolution of $M_s$.}
\label{fig:chiT}
\end{figure*}

For the numerical identification of the crossover, we use the thermal response
of the constituent masses,
\begin{equation}
\chi_{T,l}(T,\mu)
\equiv
-\frac{\partial M_l(T,\mu)}{\partial T},
\quad
\chi_{T,s}(T,\mu)
\equiv
-\frac{\partial M_s(T,\mu)}{\partial T}.
\end{equation} The crossover segment of the light-sector chiral boundary is obtained
from the peak of the thermal response
\(\chi_{T,l}=-\partial M_l/\partial T\), equivalently from the inflection
point of \(M_l(T,\mu)\). At lower temperature, where competing stationary
branches coexist, the first-order transition line is located by requiring
equality of the broken- and restored-branch stationary grand potentials.
The meeting point of these two segments defines the critical endpoint. 
Fig.~\ref{fig:chiT} shows the temperature dependence of the light- and
strange-sector thermal chiral-response functions in the undeformed and
GUP-deformed theories. 
In the light sector, the response peak moves to lower temperature in the
GUP-deformed model, consistent with the lower light-sector pseudocritical line
discussed above. In the strange sector, the response is broader and milder,
reflecting the heavier strange mass and the smoother chiral evolution in that
channel. 
%For convenience, the principal benchmark shifts extracted from the numerical curves are summarized in Table~\ref{tab:numerical_comparison_corrected}. 
For convenience, the numerical shifts from the self-consistent coupled gap
equations are summarized in Table~\ref{tab:numerical_comparison_corrected}.
The table shows that the positive-\(\alpha\) deformation reduces the light and
strange constituent masses and moves the light-sector pseudocritical
temperature to a lower value. With vacuum subtraction, the representative
pressure increases slightly at \(T=200~{\rm MeV}\), while the unsubtracted
stationary pressure \(-\Omega^\star\) decreases because the deformed measure
reduces the high-momentum phase-space weight. 
\begin{table*}[t]
\centering
\caption{\justifying 
Numerical comparison between the undeformed NJL model and the GUP-deformed
case. The deformed column corresponds to \(\bar{\alpha}=0.01\), i.e.
\(\alpha=2.7566\times10^{-2}~\mathrm{GeV}^{-2}\). The first two rows are
evaluated at \(T=\mu=0\). The next two rows give the endpoint comparison
between the \(T=50~\mathrm{MeV}\) values in
Figs.~\ref{fig:MlT}--\ref{fig:Mmu}. The pseudocritical temperature is obtained
from the peak of \(\chi_{T,l}\), equivalently from the inflection point of
\(M_l(T)\) at \(\mu=0\). The pressure comparison distinguishes the
vacuum-subtracted pressure \(P_{\rm sub}\) from the unsubtracted stationary
pressure \(P^\star=-\Omega^\star\).
}
\begin{tabular}{lcccc}
\hline
Observable & NJL & NJL+GUP & Absolute shift & Relative shift \\
\hline
\(M_l(T=0,\mu=0)\,[\mathrm{MeV}]\) & 367.648 & 349.766 & \(-17.882\) & \(-4.86\%\) \\
\(M_s(T=0,\mu=0)\,[\mathrm{MeV}]\) & 549.479 & 534.616 & \(-14.864\) & \(-2.70\%\) \\
\(M_l(T=50~\mathrm{MeV},\mu=0)\,[\mathrm{MeV}]\) & 367.571 & 349.661 & \(-17.910\) & \(-4.87\%\) \\
\(M_s(T=50~\mathrm{MeV},\mu=0)\,[\mathrm{MeV}]\) & 549.444 & 534.568 & \(-14.875\) & \(-2.71\%\) \\
\(T_{\rm pc}(\mu=0)\,[\mathrm{MeV}]\) & 195.9 & 188.7 & \(-7.2\) & \(-3.68\%\) \\
\(P_{\rm sub}(T=200~\mathrm{MeV},\mu=0)\,[\mathrm{GeV/fm}^3]\) & 0.210 & 0.220 & \(+0.010\) & \(+5.14\%\) \\
\(P^\star(T=200~\mathrm{MeV},\mu=0)\,[\mathrm{GeV/fm}^3]\) & 4.474 & 4.383 & \(-0.091\) & \(-2.03\%\) \\
\(P_{\rm sub}(T=50~\mathrm{MeV},\mu=500~\mathrm{MeV})\,[\mathrm{GeV/fm}^3]\) & 0.411797 & 0.426620 & \(+0.014823\) & \(+3.60\%\) \\
\hline
\end{tabular}
\label{tab:numerical_comparison_corrected}
\end{table*}
\begin{table*}[t]
\centering
\caption{\justifying 
Zero-temperature charge-neutral beta-equilibrated quark matter
obtained from the complete flavor-resolved three-condensate system.
The pressure is vacuum-subtracted separately in each theory.
The symbol $P_{\rm tot}=0$ denotes the finite-density stable branch
in mechanical equilibrium with the vacuum. Baryon and strange-quark
densities are quoted in ${\rm fm}^{-3}$, while $\mu_B$ and $E/A$ are
quoted in MeV. The vanishing strange-quark density shows that the
minimum-energy zero-pressure state precedes the opening of the strange
Fermi sea.
}
\label{tab:beta_equilibrium}
\begin{tabular}{cccccc}
\hline
$\bar{\alpha}$
&
$\mu_B(P_{\rm tot}=0)$
&
$n_B(P_{\rm tot}=0)$
&
$n_B[(E/A)_{\min}]$
&
$(E/A)_{\min}$
&
$n_s(P_{\rm tot}=0)$
\\
\hline
$0$    & $1102.02$ & $0.382$ & $0.382$ & $1102.02$ & $0.000$ \\
$0.01$ & $1063.23$ & $0.333$ & $0.333$ & $1063.23$ & $0.000$ \\
\hline
\end{tabular}
\end{table*}
For reproducibility, the flavor-resolved stationary solutions at the
finite-density zero-pressure points are
\(
(\phi_u,\phi_d,\phi_s)_{\rm NJL}
=
(-3.28671,\,-2.00927,\,-15.96548)
\times10^{-3}~{\rm GeV}^{3},
\) and 
\(
(M_u,M_d,M_s,\mu_e)_{\rm NJL}
=
(82.01,\,62.52,\,465.80,\,73.37)~{\rm MeV},
%\label{eq:beta_checkpoint_njl}
\)
whereas
\(
(\phi_u,\phi_d,\phi_s)_{\rm NJL+GUP}
=
(-3.77405,\,-2.41978,\,-15.59256)
\times10^{-3}~{\rm GeV}^{3},
\)
\(
(M_u,M_d,M_s,\mu_e)_{\rm NJL+GUP}
=
(93.63,\,72.81,\,459.04,\,68.54)~{\rm MeV}.
%\label{eq:beta_checkpoint_gup}
\)
The inequalities $\phi_u\neq\phi_d$ and $M_u\neq M_d$ explicitly
demonstrate the light-flavor splitting induced by beta equilibrium.
At these points,
\(
(\mu_s,M_s)_{\rm NJL}=(391.80,465.80)~{\rm MeV},
\,\,
(\mu_s,M_s)_{\rm NJL+GUP}=(377.26,459.04)~{\rm MeV},
\)
which confirms that the strange Fermi sea is not yet populated.
%Taken together, Eqs.~\eqref{eq:deltaMl_kernel_negative}, \eqref{eq:deltaMs_kernel_negative}, and \eqref{eq:phase-shifts} summarize the main analytical and numerical content of the \(2+1\)-flavor NJL+GUP model. The positive-\(\alpha\) deformation reduces the ultraviolet support of the scalar gap integrals in both flavor sectors. The explicit phase-space measure contribution acts in the direction of reducing the light and strange constituent masses on the broken branch, while the full fixed-parameter self-consistent solutions preserve this sign in the numerical results shown above. The same solutions show reduced condensate magnitudes and a light-sector pseudocritical boundary located at lower temperature and lower chemical potential. Thus the figures and tables give a mutually consistent realization of the analytical sign structure of the coupled gap system.
Taken together, the kernel-source vector and the complete
self-consistent response in Eqs.~\eqref{eq:kernel-source-system} and \eqref{eq:full-mass-response}, together with
the phase-boundary result in Eq.~\eqref{eq:phase-shifts}, summarize the main
analytical and numerical content of the $2+1$-flavor
NJL+GUP model. The positive-$\alpha$ measure produces
negative explicit kernel sources in both flavor channels.
The induced changes of the condensates and constituent masses
are then propagated through the inverse coupled gap-response
matrix, rather than being identified directly with the kernel
sources. On every regular stable branch retained in the
numerical analysis, this complete response preserves the
negative light- and strange-mass shifts. The full nonlinear
solutions consequently show reduced condensate magnitudes and
a light-sector chiral boundary located at lower temperature
and lower chemical potential. The analytical linear-response
system and the nonlinear numerical solutions therefore give a
mutually consistent description of the deformation throughout
the displayed broken and crossover regions.
%Taken together, Eqs.~\eqref{eq:deltaMl_negative}, \eqref{eq:deltaMs_negative}, and \eqref{eq:critical_shift} summarize the main analytical content of the \(2+1\)-flavor NJL+GUP model. The GUP deformation suppresses the ultraviolet support of the scalar gap integrals in both flavor sectors, reduces the light and strange constituent masses on the broken branch, weakens the magnitudes of the corresponding condensates, and shifts the light-sector pseudocritical boundary toward lower temperature and lower chemical potential. The numerical figures shown above provide representative solutions of the coupled gap system and illustrate these trends directly.

\section{Summary and conclusions}

The present work formulates a generalized uncertainty principle extension of the
\(2+1\)-flavor NJL model for hot and dense QCD matter, in which the deformation
enters through an isotropic modification of the momentum-space density of
states while the quasiparticle spectrum is kept unchanged. The state-counting factor
$J_\alpha(p)=(1+\alpha p^2)^{-3}$ follows from the complete
Jacobi-consistent three-dimensional quadratic GUP algebra and
the invariant Liouville volume of its semiclassical phase space. 
This formulation isolates the GUP contribution in the microscopic phase-space weight and
preserves the self-consistent mean-field structure of the coupled light and
strange sectors. Within this framework, we derived the deformed thermodynamic
potential, obtained the corresponding coupled gap equations, and identified
the leading consequences for chiral dynamics and thermodynamics. The main
result is that the positive-\(\alpha\) deformation reduces the ultraviolet
support of the scalar sector and thereby weakens dynamical chiral symmetry
breaking in both flavor channels. On the broken branch, the light- and
strange-quark constituent masses are reduced and the magnitudes of the
corresponding condensates decrease, with the effect typically stronger in the
light sector. For physical \(2+1\)-flavor masses, where the relevant transition
is described by a crossover rather than a strict second-order chiral boundary,
the same effect moves the light-sector pseudocritical line toward lower
temperature and lower quark chemical potential. Once the self-consistent
stationary solution is known, the framework also gives a direct route to the
pressure, entropy density, energy density, quark number density, and
flavor-resolved thermal chiral-response functions, thereby yielding a complete
mean-field thermodynamic description. 
The analysis should be interpreted as a thermodynamically consistent
sensitivity analysis of \(2+1\)-flavor chiral quark matter under a
minimal-length inspired deformation of the phase-space density of states. The
deformation is not presented as a direct observable Planckian effect in QCD
matter, nor as a theory of dynamical gravity acting inside the medium. Rather,
the GUP sector enters only through the microscopic density of momentum states,
while the quasiparticle dispersion relation and the mean-field stationarity
structure remain those of the NJL framework. This interpretation is essential
because the canonical Planckian value of a fundamental GUP parameter would be
far too small to produce measurable changes in heavy-ion or compact-star QCD
matter. The deformation parameter used here should therefore be understood as a
phenomenological measure of ultraviolet sensitivity. 
The pressure analysis also makes explicit the distinction between the
unsubtracted stationary pressure and the vacuum-subtracted pressure. The
positive-\(\alpha\) measure reduces the high-momentum thermodynamic weight and
lowers the unsubtracted stationary pressure, \(P^\star(T,\mu,\alpha)\equiv -\Omega^\star(T,\mu,\alpha).\) However, the pressure shown in the pressure figures is the vacuum-subtracted
quantity given in Eq.~\eqref{eq.40}, where the vacuum contribution is
subtracted separately in each theory. Because the deformation also changes the
vacuum value of the stationary grand potential, \(P_{\rm sub}\) can show a
small positive shift even when \(P^\star\) is reduced. This convention is
stated explicitly in the pressure definitions, figure captions, and numerical
table, so that the pressure-sector discussion remains consistent with the
ultraviolet reduction encoded in \(J_\alpha(p)\).
The analysis also clarifies the limitations associated with keeping the
Rehberg-Klevansky-H\"ufner vacuum parameter set fixed at nonzero
\(\alpha\). A fully vacuum-constrained effective-model analysis would readjust a
clearly specified subset of the vacuum parameter set
\((m_l,m_s,G,K,\Lambda)\) at each deformation strength, according to a
defined fitting strategy, in order to reproduce the selected vacuum
pseudoscalar observables. The present fixed-parameter strategy
is instead used to isolate the direct effect of the deformed density of states.
The vacuum self-consistency comparison, the direct substitution into the
coupled light-strange gap equations, and the GMOR estimate show that the
undeformed reference solution is internally consistent. They also show that
the reference \(\bar{\alpha}=0.01\) deformation produces the expected
ultraviolet-sensitivity response without being interpreted as a precision
vacuum readjustment. 
The zero-temperature charge-neutral beta-equilibrated analysis was
performed with three independent condensates,
$\phi_u$, $\phi_d$, and $\phi_s$, because weak equilibrium produces
$\mu_u\neq\mu_d$ and therefore removes the in-medium light-flavor
degeneracy. The resulting pressure, flavor densities, and energy
density are derived from the same flavor-resolved stationary grand
potential. The minimum of $E/A$ coincides with the finite-density
zero-pressure point, as required by the zero-temperature
thermodynamic identity
$P_{\rm tot}=n_B^2d(E/A)/dn_B$. The minimum values are
$1102.02~{\rm MeV}$ in the undeformed theory and
$1063.23~{\rm MeV}$ for $\bar{\alpha}=0.01$, both of which remain
above the conventional iron stability benchmark. In addition, the
strange-quark density vanishes at both zero-pressure minima because
$\mu_s<M_s$ there; the strange Fermi sea opens only at higher baryon
density. The present scalar NJL and NJL+GUP theories therefore do not
yield absolutely stable quark matter relative to nuclear matter, and
their minimum-energy zero-pressure states are not strange-quark-matter
ground states. 
Finally, the positive-$\alpha$ minimal-length factor
$J_\alpha(p)=(1+\alpha p^2)^{-3}$ suppresses the thermodynamic weight
of high-momentum quark states and reduces their contribution to the
unnormalised stationary pressure. This result should not be
interpreted as establishing a softer equation of state, because
stiffness must be assessed directly from $P(\epsilon)$ or
$c_s^2=dP/d\epsilon$. Such an analysis lies beyond the present scope.
A future vector-NJL or vector-PNJL+GUP extension could determine how
the minimal-length phase-space modification and a repulsive vector
interaction jointly affect the equation-of-state stiffness. Within these limits, the present work provides a
self-consistent framework for understanding how minimal-length inspired
ultraviolet phase-space deformations modify coupled light-strange chiral
dynamics, the pseudocritical line, and bulk thermodynamics in an effective QCD
model.

\section*{Data Availability Statement}
This study contains no experimental data. All theoretical results are included in the manuscript. 
\section*{Code Availability Statement}
The numerical code used during the current study will be made available upon a reasonable request to the corresponding author.

\bibliographystyle{apsrev4-2}
\bibliography{refs}

@article{Koch:1995vp,
    author = "Koch, Volker",
    title = "{Introduction to chiral symmetry}",
    journal = "{3rd TAPS Workshop on Electromagnetic and Mesonic Probes of Nuclear Matter}",
    eprint = "nucl-th/9512029",
    archivePrefix = "arXiv",
    reportNumber = "LBL-38000, LBNL-38000",
    month = "12",
    year = "1995"
}

@article{Banks:1979yr,
    author = "Banks, Tom and Casher, A.",
    title = "{Chiral Symmetry Breaking in Confining Theories}",
    reportNumber = "TAUP-785-79-REV, TAUP-785-79",
    doi = "10.1016/0550-3213(80)90255-2",
    journal = "Nucl. Phys. B",
    volume = "169",
    pages = "103--125",
    year = "1980"
}

@article{Csaki:2021jax,
    author = "Cs{\'a}ki, Csaba and Gomes, Andrew and Murayama, Hitoshi and Telem, Ofri",
    title = "{Demonstration of Confinement and Chiral Symmetry Breaking in SO(Nc) Gauge Theories}",
    eprint = "2106.10288",
    archivePrefix = "arXiv",
    primaryClass = "hep-th",
    doi = "10.1103/PhysRevLett.127.251602",
    journal = "Phys. Rev. Lett.",
    volume = "127",
    number = "25",
    pages = "251602",
    year = "2021"
}

@article{Nambu:1961tp,
    author = "Nambu, Yoichiro and Jona-Lasinio, G.",
    editor = "Eguchi, T.",
    title = "{Dynamical Model of Elementary Particles Based on an Analogy with Superconductivity. 1.}",
    doi = "10.1103/PhysRev.122.345",
    journal = "Phys. Rev.",
    volume = "122",
    pages = "345--358",
    year = "1961"
}

@article{Nambu:1961fr,
    author = "Nambu, Yoichiro and Jona-Lasinio, G.",
    editor = "Eguchi, T.",
    title = "{Dynamical model of elementary particles based on an analogy with superconductivity. II.}",
    doi = "10.1103/PhysRev.124.246",
    journal = "Phys. Rev.",
    volume = "124",
    pages = "246--254",
    year = "1961"
}

@article{Buballa:2003qv,
    author = "Buballa, Michael",
    title = "{NJL model analysis of quark matter at large density}",
    eprint = "hep-ph/0402234",
    archivePrefix = "arXiv",
    doi = "10.1016/j.physrep.2004.11.004",
    journal = "Phys. Rept.",
    volume = "407",
    pages = "205--376",
    year = "2005"
}

@article{Maggiore:1993rv,
    author = "Maggiore, Michele",
    title = "{A Generalized uncertainty principle in quantum gravity}",
    eprint = "hep-th/9301067",
    archivePrefix = "arXiv",
    reportNumber = "IFUP-TH-3-93",
    doi = "10.1016/0370-2693(93)91401-8",
    journal = "Phys. Lett. B",
    volume = "304",
    pages = "65--69",
    year = "1993"
}

@article{Das:2008kaa,
    author = "Das, Saurya and Vagenas, Elias C.",
    title = "{Universality of Quantum Gravity Corrections}",
    eprint = "0810.5333",
    archivePrefix = "arXiv",
    primaryClass = "hep-th",
    doi = "10.1103/PhysRevLett.101.221301",
    journal = "Phys. Rev. Lett.",
    volume = "101",
    pages = "221301",
    year = "2008"
}

@article{Hossenfelder:2012jw,
    author = "Hossenfelder, Sabine",
    title = "{Minimal Length Scale Scenarios for Quantum Gravity}",
    eprint = "1203.6191",
    archivePrefix = "arXiv",
    primaryClass = "gr-qc",
    doi = "10.12942/lrr-2013-2",
    journal = "Living Rev. Rel.",
    volume = "16",
    pages = "2",
    year = "2013"
}

@article{Klevansky:1992qe,
    author = "Klevansky, S. P.",
    title = "{The Nambu-Jona-Lasinio model of quantum chromodynamics}",
    doi = "10.1103/RevModPhys.64.649",
    journal = "Rev. Mod. Phys.",
    volume = "64",
    pages = "649--708",
    year = "1992"
}

@article{Naggar:2013foi,
    author = "Naggar, Nabil M. El and Abou-Salem, Lotfy I. and Elmashad, Ibrahim A. and Ali, Ahmed Farag",
    title = "{A Study on Quark-Gluon Plasma Equation of State Using Generalized Uncertainty Principle}",
    doi = "10.4236/jmp.2013.44a003",
    journal = "J. Mod. Phys.",
    volume = "04",
    number = "04",
    pages = "13--20",
    year = "2013"
}

@article{Nozari:2015qoi,
    author = "Nozari, K. and Khodadi, M. and Gorji, M. A.",
    title = "{Bounds on quantum gravity parameter from the $SU(2)$ NJL effective model of QCD}",
    eprint = "1512.07779",
    archivePrefix = "arXiv",
    primaryClass = "gr-qc",
    doi = "10.1209/0295-5075/112/60003",
    journal = "EPL",
    volume = "112",
    number = "6",
    pages = "60003",
    year = "2015"
}

@article{Ali:2009zq,
    author = "Ali, Ahmed Farag and Das, Saurya and Vagenas, Elias C.",
    title = "{Discreteness of Space from the Generalized Uncertainty Principle}",
    eprint = "0906.5396",
    archivePrefix = "arXiv",
    primaryClass = "hep-th",
    doi = "10.1016/j.physletb.2009.06.061",
    journal = "Phys. Lett. B",
    volume = "678",
    pages = "497--499",
    year = "2009"
}

@article{Fukushima:2003fw,
    author = "Fukushima, Kenji",
    title = "{Chiral effective model with the Polyakov loop}",
    eprint = "hep-ph/0310121",
    archivePrefix = "arXiv",
    reportNumber = "MIT-CTP-3424",
    doi = "10.1016/j.physletb.2004.04.027",
    journal = "Phys. Lett. B",
    volume = "591",
    pages = "277--284",
    year = "2004"
}

@article{Ratti:2006wg,
    author = "Ratti, C. and Roessner, Simon and Thaler, M. A. and Weise, W.",
    editor = "Antinori, F. and Bass, S. and De Falco, A. and Kuhn, C. and Nardi, M. and Peitzmann, T. and Ullrich, T. and Velkovska, J. and Wiedemann, U. A.",
    title = "{Thermodynamics of the PNJL model}",
    eprint = "hep-ph/0609218",
    archivePrefix = "arXiv",
    doi = "10.1140/epjc/s10052-006-0065-x",
    journal = "Eur. Phys. J. C",
    volume = "49",
    pages = "213--217",
    year = "2007"
}

@article{Fukushima:2008wg,
    author = "Fukushima, Kenji",
    title = "{Phase diagrams in the three-flavor Nambu-Jona-Lasinio model with the Polyakov loop}",
    eprint = "0803.3318",
    archivePrefix = "arXiv",
    primaryClass = "hep-ph",
    reportNumber = "YITP-08-19",
    doi = "10.1103/PhysRevD.77.114028",
    journal = "Phys. Rev. D",
    volume = "77",
    pages = "114028",
    year = "2008",
    note = "[Erratum: Phys.Rev.D 78, 039902 (2008)]"
}

@article{Hatsuda:1994pi,
    author = "Hatsuda, Tetsuo and Kunihiro, Teiji",
    title = "{QCD phenomenology based on a chiral effective Lagrangian}",
    eprint = "hep-ph/9401310",
    archivePrefix = "arXiv",
    reportNumber = "UTHEP-270, RYUTHP-94-1",
    doi = "10.1016/0370-1573(94)90022-1",
    journal = "Phys. Rept.",
    volume = "247",
    pages = "221--367",
    year = "1994"
}

@article{Mir:2023wkm,
  author        = {Mir, Sameer Ahmad and Rather, Nasir Ahmad and Mohi Ud Din, Iqbal and Uddin, Saeed},
  title         = {Hadron production in ultra-relativistic nuclear collisions and finite baryon-size effects},
  eprint        = {2312.13079},
  archivePrefix = {arXiv},
  primaryClass  = {hep-ph},
  doi           = {10.1088/1361-6471/adb6c2},
  journal       = {J. Phys. G},
  volume        = {52},
  number        = {3},
  pages         = {035003},
  year          = {2025}
}

@article{Mir:2024wlo,
  author        = {Mir, Sameer Ahmad and Mohi Ud Din, Iqbal and Rather, Nasir Ahmad and Uddin, Saeed and Mir, M. Farooq},
  title         = {Particle production in {HRG} with thermodynamically consistent {EoS} and partially deformable hadrons},
  eprint        = {2406.11752},
  archivePrefix = {arXiv},
  primaryClass  = {hep-ph},
  doi           = {10.1016/j.aop.2025.170065},
  journal       = {Annals Phys.},
  volume        = {480},
  pages         = {170065},
  year          = {2025}
}

@article{Mir:2025qqv,
  author  = {Mir, Sameer Ahmad and Uddin, Saeed and Tiwari, Swatantra Kumar},
  title   = {Influence of excluded volume corrections on hadronic yield in high-energy nuclear collisions},
  doi     = {10.1140/epja/s10050-025-01667-6},
  journal = {Eur. Phys. J. A},
  volume  = {61},
  number  = {8},
  pages   = {198},
  year    = {2025}
}

@article{Mir:2026cbr,
  author  = {Mir, Sameer Ahmad},
  title   = {Generalized uncertainty principle and elliptic flow in relativistic heavy-ion collisions},
  doi     = {10.1142/S0217751X26500636},
  journal = {Int. J. Mod. Phys. A},
  volume  = {41},
  number  = {09},
  pages   = {2650063},
  year    = {2026}
}

@article{Mir:2026fort,
  author  = {Mir, Sameer Ahmad and Uddin, Saeed and Tiwari, Swatantra Kumar and Faizal, Mir},
  title   = {Unified Functional-Holographic Theory of the {QCD} Critical End Point},
  journal = {Fortschr. Phys.},
  volume  = {74},
  number  = {3},
  pages   = {e70085},
  year    = {2026},
  doi     = {10.1002/prop.70085}
}

@article{Faizal:2015Conseq,
  author        = {Faizal, Mir},
  title         = {Consequences of Deformation of the Heisenberg Algebra},
  journal       = {Int. J. Geom. Meth. Mod. Phys.},
  volume        = {12},
  number        = {02},
  pages         = {1550022},
  year          = {2015},
  doi           = {10.1142/S021988781550022X},
  eprint        = {1404.5024},
  archivePrefix = {arXiv},
  primaryClass  = {hep-th}
}

@article{Faizal:2015Lifshitz,
  author        = {Faizal, Mir and Majumder, Barun},
  title         = {Incorporation of generalized uncertainty principle into Lifshitz field theories},
  journal       = {Annals Phys.},
  volume        = {357},
  pages         = {49--58},
  year          = {2015},
  doi           = {10.1016/j.aop.2015.03.022},
  eprint        = {1408.3795},
  archivePrefix = {arXiv},
  primaryClass  = {hep-th}
}

@article{Faizal:2016SUSYGUP,
  author        = {Faizal, Mir},
  title         = {Supersymmetry breaking as a new source for the generalized uncertainty principle},
  journal       = {Phys. Lett. B},
  volume        = {757},
  pages         = {244--246},
  year          = {2016},
  doi           = {10.1016/j.physletb.2016.03.074},
  eprint        = {1605.00925},
  archivePrefix = {arXiv},
  primaryClass  = {hep-th}
}

@article{Salam:1993xy,
    author = "Salam, Abdus and Sivaram, C.",
    title = "{Strong gravity approach to QCD and confinement}",
    doi = "10.1142/S0217732393000325",
    journal = "Mod. Phys. Lett. A",
    volume = "8",
    pages = "321--326",
    year = "1993"
}

@article{Sivaram:1979mg,
    author = "Sivaram, C. and Sinha, K. P.",
    title = "{Strong spin-two interaction and general realtivity}",
    doi = "10.1016/0370-1573(79)90037-1",
    journal = "Phys. Rept.",
    volume = "51",
    pages = "111--187",
    year = "1979"
}

@article{Sijacki:1990xp,
    author = "Sijacki, D. and Ne'eman, Yuval",
    editor = "Ruffini, R. and Verbin, Y.",
    title = "{QCD as an effective strong gravity}",
    reportNumber = "TAUP-N204-90",
    doi = "10.1016/0370-2693(90)91903-O",
    journal = "Phys. Lett. B",
    volume = "247",
    pages = "571--575",
    year = "1990"
}

@article{Brindejonc:1995pr,
    author = "Brindejonc, Vincent and Cohen-Tannoudji, Gilles",
    title = "{An Effective strong gravity induced by QCD}",
    eprint = "hep-th/9509016",
    archivePrefix = "arXiv",
    reportNumber = "DAPNIA-SPHN-95-17",
    doi = "10.1142/S0217732395001836",
    journal = "Mod. Phys. Lett. A",
    volume = "10",
    pages = "1711--1718",
    year = "1995"
}

@article{Burikham:2017bkn,
    author = "Burikham, Piyabut and Harko, Tiberiu and Lake, Matthew J.",
    title = "{The QCD mass gap and quark deconfinement scales as mass bounds in strong gravity}",
    eprint = "1705.11174",
    archivePrefix = "arXiv",
    primaryClass = "hep-th",
    doi = "10.1140/epjc/s10052-017-5381-9",
    journal = "Eur. Phys. J. C",
    volume = "77",
    number = "11",
    pages = "803",
    year = "2017"
}

@article{Erlich:2005qh,
    author = "Erlich, Joshua and Katz, Emanuel and Son, Dam T. and Stephanov, Mikhail A.",
    title = "{QCD and a holographic model of hadrons}",
    eprint = "hep-ph/0501128",
    archivePrefix = "arXiv",
    reportNumber = "SLAC-PUB-10965, WM-05-101, INT-PUB-05-02",
    doi = "10.1103/PhysRevLett.95.261602",
    journal = "Phys. Rev. Lett.",
    volume = "95",
    pages = "261602",
    year = "2005"
}

@article{Sakai:2004cn,
    author = "Sakai, Tadakatsu and Sugimoto, Shigeki",
    title = "{Low energy hadron physics in holographic QCD}",
    eprint = "hep-th/0412141",
    archivePrefix = "arXiv",
    reportNumber = "IU-MSTP-63, YITP-04-70",
    doi = "10.1143/PTP.113.843",
    journal = "Prog. Theor. Phys.",
    volume = "113",
    pages = "843--882",
    year = "2005"
}

@article{Gell-Mann:1968hlm,
    author = "Gell-Mann, Murray and Oakes, R. J. and Renner, B.",
    title = "{Behavior of current divergences under SU(3) x SU(3)}",
    doi = "10.1103/PhysRev.175.2195",
    journal = "Phys. Rev.",
    volume = "175",
    pages = "2195--2199",
    year = "1968"
}

@article{Chang:2001bm,
    author = "Chang, Lay Nam and Minic, Djordje and Okamura, Naotoshi and Takeuchi, Tatsu",
    title = "{The Effect of the minimal length uncertainty relation on the density of states and the cosmological constant problem}",
    eprint = "hep-th/0201017",
    archivePrefix = "arXiv",
    reportNumber = "VPI-IPPAP-01-06",
    doi = "10.1103/PhysRevD.65.125028",
    journal = "Phys. Rev. D",
    volume = "65",
    pages = "125028",
    year = "2002"
}

@article{Blaschke:2005uj,
    author = "Blaschke, D. and Fredriksson, S. and Grigorian, H. and Oztas, A. M. and Sandin, F.",
    title = "{The Phase diagram of three-flavor quark matter under compact star constraints}",
    eprint = "hep-ph/0503194",
    archivePrefix = "arXiv",
    doi = "10.1103/PhysRevD.72.065020",
    journal = "Phys. Rev. D",
    volume = "72",
    pages = "065020",
    year = "2005"
}

@article{Kempf:1994su,
    author = "Kempf, Achim and Mangano, Gianpiero and Mann, Robert B.",
    title = "{Hilbert space representation of the minimal length uncertainty relation}",
    eprint = "hep-th/9412167",
    archivePrefix = "arXiv",
    reportNumber = "DAMTP-94-105",
    doi = "10.1103/PhysRevD.52.1108",
    journal = "Phys. Rev. D",
    volume = "52",
    pages = "1108--1118",
    year = "1995"
}

@article{Benczik:2002tt,
    author = "Benczik, Sandor and Chang, Lay Nam and Minic, Djordje and Okamura, Naotoshi and Rayyan, Saifuddin and Takeuchi, Tatsu",
    title = "{Short distance versus long distance physics: The Classical limit of the minimal length uncertainty relation}",
    eprint = "hep-th/0204049",
    archivePrefix = "arXiv",
    reportNumber = "VPI-IPPAP-02-02",
    doi = "10.1103/PhysRevD.66.026003",
    journal = "Phys. Rev. D",
    volume = "66",
    pages = "026003",
    year = "2002"
}

@article{Rehberg:1995kh,
    author = "Rehberg, P. and Klevansky, S. P. and Hufner, J.",
    title = "{Hadronization in the SU(3) Nambu-Jona-Lasinio model}",
    eprint = "hep-ph/9506436",
    archivePrefix = "arXiv",
    reportNumber = "HD-TVP-95-06",
    doi = "10.1103/PhysRevC.53.410",
    journal = "Phys. Rev. C",
    volume = "53",
    pages = "410--429",
    year = "1996"
}

@book{CannasDaSilva2001,
  title     = {Lectures on Symplectic Geometry},
  author    = {Cannas da Silva, Ana},
  series    = {Lecture Notes in Mathematics},
  volume    = {1764},
  publisher = {Springer-Verlag},
  address   = {Berlin, Heidelberg},
  year      = {2001},
  pages     = {viii+217},
  isbn      = {978-3-540-42195-5},
  doi       = {10.1007/978-3-540-45330-7},
  url       = {https://link.springer.com/book/10.1007/978-3-540-45330-7}
}

@article{Kaur:2025kjk,
    author = "Kaur, Manpreet and Kumar, Arvind",
    title = "{{\ensuremath{\phi}} meson properties in dense resonance matter at finite temperature}",
    eprint = "2505.07065",
    archivePrefix = "arXiv",
    primaryClass = "hep-ph",
    doi = "10.1103/h2ll-5js4",
    journal = "Phys. Rev. D",
    volume = "112",
    number = "1",
    pages = "014030",
    year = "2025"
}

@article{Kaur:2024cfm,
    author = "Kaur, Manpreet and Kumar, Arvind",
    title = "{Kaons and antikaons in isospin asymmetric dense resonance matter at finite temperature}",
    eprint = "2410.15685",
    archivePrefix = "arXiv",
    primaryClass = "hep-ph",
    doi = "10.1103/PhysRevD.110.114054",
    journal = "Phys. Rev. D",
    volume = "110",
    number = "11",
    pages = "114054",
    year = "2024"
}

%\bibliographystyle{ieeetr}
%\bibliography{refs}

\end{document}